\documentclass[a4paper,11pt]{article}
\pdfoutput=1
\usepackage{subcaption}
\usepackage{jcappub}
\usepackage{xcolor}
\usepackage{amsmath}
\usepackage{booktabs}

\title{Cosmic Strings as Dynamical Dark Energy: Novel Constraints}

\author[a,b]{Hanyu Cheng}
\author[b]{\!, Eleonora Di Valentino}
\author[c,d]{\!, and Luca Visinelli}

\affiliation[a]{Tsung-Dao Lee Institute \& School of Physics and Astronomy, Shanghai Jiao Tong University, Shanghai 201210, China}
\affiliation[b]{School of Mathematics and Statistics, University of Sheffield, Hounsfield Road, Sheffield S3 7RH, United Kingdom}
\affiliation[c]{Dipartimento di Fisica ``E.R.\ Caianiello'', Universit\`a degli Studi di Salerno,\\ Via Giovanni Paolo II, 132 - 84084 Fisciano (SA), Italy}
\affiliation[d]{Istituto Nazionale di Fisica Nucleare - Gruppo Collegato di Salerno - Sezione di Napoli,\\ Via Giovanni Paolo II, 132 - 84084 Fisciano (SA), Italy}

\emailAdd{hcheng19@sheffield.ac.uk}
\emailAdd{e.divalentino@sheffield.ac.uk}
\emailAdd{lvisinelli@unisa.it}

\abstract{Cosmic strings, topological defects predicted by high-energy theories, may contribute to the late-time expansion of the Universe, effectively mimicking dynamical dark energy. We investigate four phenomenological extensions of the $\Lambda$CDM model involving a residual string network: (i) a non-relativistic component with positive energy density (Model~1), (ii) a velocity-dependent extension (Model~2), (iii) a non-relativistic string network with energy density allowed to take both positive and negative values (Model~3), and (iv) a general scenario with free energy and velocity parameters (Model~4). These models are constrained using \textit{Planck} CMB data, SDSS or DESI baryon acoustic oscillations, and Type Ia supernovae observations. Models~1 and~2 yield strong upper bounds on the string density, for example, $\Omega_{\mathrm{s}} < 0.00901$ at 95\% CL from the CMB+DESI+DESY5 combination for Model~2, and mildly shift the inferred value of $H_0$ upward, though they are not favored by Bayesian evidence. For the same combination, the bulk velocity is bound as $v_{\mathrm{s}} < 0.569$. Models~3 and~4 exhibit a consistent preference for slightly negative values of $\Omega_{\mathrm{s}}$, with CMB-only data yielding $\Omega_{\mathrm{s}} = -0.038^{+0.029}_{-0.022}$ and $v_{\mathrm{s}}< 0.574$ in Model~4, and a best-fit improvement of $\Delta \chi^2 = -6.07$. However, these improvements are not sufficient to overcome the Occam penalty, and the Bayesian evidence continues to favor $\Lambda$CDM. These findings demonstrate the power of current data to constrain exotic energy components and encourage further exploration of string-inspired extensions to $\Lambda$CDM, particularly those involving negative-tension networks.}

\begin{document}
\maketitle
\flushbottom

%++++++++++++++++++++++++++++++++
\section{Introduction}
\label{sec:Introduction}

Cosmic strings are one-dimensional topological defects that may have formed during symmetry-breaking phase transitions in the early universe, as predicted by various grand unified theories (GUTs) and extensions of the Standard Model~\cite{Kibble:1976sj, Vilenkin:1981bx, Vilenkin:1984ib, Witten:1984eb, Vilenkin:2000jqa}. These objects are characterized by a large energy per unit length, or tension $\mu$, and exhibit relativistic dynamics due to their high tension. The gravitational effects of cosmic strings are typically parametrized by the dimensionless quantity $G\mu$, where $G$ is Newton's constant. Observational efforts aim to constrain this parameter through a variety of astrophysical and cosmological probes. A network of cosmic strings is expected to evolve towards a scaling regime, where the statistical properties of the network remain constant relative to the Hubble scale~\cite{Bennett:1989yp, Albrecht:1989mk}. This evolution includes the production of closed loops through string intercommutations. These loops oscillate and lose energy predominantly via gravitational radiation, leading to a stochastic gravitational wave background (SGWB), as well as occasional bursts from localized features such as cusps and kinks~\cite{Damour:2000wa, Siemens:2006yp}.

Confronting experimental CMB data with numerical simulations of cosmic string networks leads to an upper bound on the string tension of $G\mu \lesssim 10^{-7}$~\cite{Planck:2013mgr, Henrot-Versille:2014jua, Lazanu:2014eya, Lizarraga:2016onn}. These constraints arise from the imprint of long strings on the CMB anisotropy power spectrum, which rules out models where strings dominate structure formation. In the realm of GW searches, the LIGO and Virgo collaborations have searched for burst-like signals from cosmic string cusps and kinks. The null results from Advanced LIGO's first observing run (O1) have been used to place upper bounds on $G\mu$ for a variety of cosmic string loop models, with typical constraints reaching $G\mu \lesssim 10^{-11}\textrm{--}10^{-8}$, depending on the assumed parameters and reconnection probabilities~\cite{LIGOScientific:2017ikf}. The next generation of gravitational wave detectors is expected to further enhance sensitivity to cosmic string signals. Space-based interferometers such as LISA will be sensitive to lower-frequency gravitational waves, making them well-suited to probing cosmic string loops formed in the early universe~\cite{Auclair:2019wcv, LISACosmologyWorkingGroup:2022jok}. Ground-based third-generation detectors like the Einstein Telescope (ET) and Cosmic Explorer (CE) will extend sensitivity to a broader frequency range, potentially accessing string tensions as low as $G\mu \sim 10^{-16}$~\cite{Punturo:2010zz, Sathyaprakash:2012jk, Maggiore:2019uih}. Pulsar timing arrays (PTAs), including the North American Nanohertz Observatory for Gravitational Waves (NANOGrav) and the European Pulsar Timing Array (EPTA) collaborations, also provide complementary constraints at nanohertz frequencies and may already be hinting at a stochastic background consistent with cosmic string origins~\cite{NANOGrav:2023hvm, EPTA:2023xxk}. These future observations will enable a multi-band approach to test the existence of cosmic strings across a vast range of energy and length scales~\cite{Cui:2017ufi, Cui:2018rwi, Chang:2019mza, Gouttenoire:2021jhk}. Non-standard cosmic string scenarios, in which the string network radiates energy not only through gravitational waves but also via light gauge bosons, can be probed through multimessenger observations. In these models, the coupling between the strings and hidden-sector fields leads to additional radiation channels, giving rise to a gravitational wave spectrum and other novel signatures~\cite{Dror:2018pdh, Ramberg:2019dgi, Long:2019lwl, Ramberg:2020oct, Cheng:2024axj}.

It remains a possibility that a subdominant fraction of cosmic strings, with energy density $\rho_s$, has survived until the present epoch and contributes to the overall expansion rate of the Universe. While this scenario has been previously considered~\cite{Battye:2006mb, Pogosian:2008am, Lizarraga:2016onn, Silva:2023diq, Rybak:2024djq}, it has not been systematically excluded. The release of recent cosmological datasets with improved control over systematics, such as those from \textit{Planck} and large-scale structure surveys, offers an opportunity to revisit and constrain this hypothesis with greater precision. If present, cosmic strings would constitute an exotic form of energy density beyond the standard dark matter and dark energy components, whose presence is well established by multiple cosmological probes. The inclusion of strings could modify the late-time expansion history and the growth of structure, potentially alleviating persistent cosmological tensions such as the Hubble tension or the $S_8$ discrepancy~\cite{DiValentino:2021izs, Abdalla:2022yfr, DiValentino:2025sru}. To enable a fully general analysis unconstrained by priors on the sign of the energy density, we allow the cosmic string density parameter to assume both positive and negative values. This approach is inspired by previous studies of negative energy density components in cosmology~\cite{Nemiroff:2014gea, Visinelli:2019qqu}, and is supported by theoretical constructions in which cosmic strings with negative energy density arise~\cite{Aros:1997kk}.

Geometries involving negative-tension cosmic strings, corresponding to a negative mass per unit length, have been proposed in the literature~\cite{Visser:1989kh, Sushkov:2005kj, Bronnikov:2013coa}. While such configurations violate standard energy conditions and are typically regarded as unphysical within general relativity, they open intriguing possibilities in the context of exotic matter and non-trivial spacetime topologies. A particularly compelling proposal is that negative-mass strings could stabilize traversable wormholes~\cite{Visser:1989kh, Visser:2003yf, Morris:1988cz}. If such strings wrapped around a wormhole throat in the early Universe, they could provide the negative energy density needed to prevent collapse, potentially allowing the wormhole to survive to the present day. Although speculative, such scenarios motivate continued efforts to test the limits of classical energy conditions. Violations of the weak and null energy conditions are being probed through cosmological observations~\cite{Barcelo:2002bv, Rubakov:2014jja}, semi-classical quantum field theory~\cite{Ford:1994bj, Fewster:2005gp}, and high-precision laboratory experiments~\cite{Davies:2002bg, Adelberger:2009zz, Burrage:2017qrf}. Negative-energy components can naturally emerge in string-inspired setups. For example, in some brane-world scenarios, negative-tension branes arise as consistent solutions in flux compactifications~\cite{Kachru:2003aw}. Additionally, effective negative energy densities may result from ghost condensates~\cite{Arkani-Hamed:2003pdi} or modified couplings between strings and hidden-sector fields~\cite{Carroll:2003st}, both capable of violating classical energy conditions. These constructions provide further motivation for exploring negative $\Omega_{\mathrm{s}}$ in a cosmological context.

This paper is structured as follows. In section~\ref{model}, we introduce the theoretical framework and model definitions. Section~\ref{sec:methods} describes the methodology and datasets. Section~\ref{results} presents the results for models with positive energy density, while section~\ref{discussion} discusses the case where negative energy contributions are allowed. Conclusions are summarized in section~\ref{conclusions}.

\section{Model}\label{model}

We describe the relation between the pressure and energy density of a cosmic string network as $p_s = w_s \rho_s$~\cite{Vachaspati:1986cc}, where the parameter $w_s$ depends on the root-mean squared velocity of long strings $v_{\mathrm{s}}$ through the relation~\cite{Kumar:2012dv, Martins:1996jp, Martins:2000cs}
\begin{equation}
    w_s = \frac{2v_{\mathrm{s}}^2 - 1}{3}.
\end{equation}
This effective equation of state arises from averaging over the dynamics of a network of long strings. The parameter $w_s$ varies within the range $-1/3 \leq w_s \leq 1/3$, reflecting the transition from non-relativistic to ultra-relativistic motion. For example, ultra-relativistic strings with $v_{\mathrm{s}}^2 \approx 1$ behave like a radiation fluid with $w_s = 1/3$. Conversely, slowly moving strings with $v_{\mathrm{s}}^2 \approx 0$ yield $w_s = -1/3$, corresponding to a network whose energy density dilutes more slowly than radiation or matter and can potentially dominate the energy density of the Universe. The evolution of the string energy density for a network of non-interacting and non-intercommutating strings~\cite{Spergel:1996ai, Sousa:2011iu,Sousa:2009is} is governed by the continuity equation
\begin{equation}
    \frac{{\rm d}\rho_s}{{\rm d}t} + 3H(1+w_s)\rho_s = 0,
    \label{eq:freestrings}
\end{equation}
which has the solution $\rho_s \propto a^{-3(1+w_s)}$, where $a$ is the scale factor. In the limiting case $w_s = -1/3$, the evolution simplifies to $\dot{\rho}_s + 2H \rho_s = 0$, where the factor of two arises from the combination of dilution due to expansion and the string stretching term, such that the energy per unit length remains constant under Hubble expansion. In this case, the energy density evolves as $\rho_s \propto a^{-2}$, meaning that a free-string network with $w_s = -1/3$ would rapidly dominate the expansion. However, realistic string networks do not evolve freely. Numerical simulations and analytic modeling within the velocity-dependent one-scale framework~\cite{Martins:1996jp, Martins:2000cs} show that long strings interact through intercommutation and self-intersections, leading to the formation of loops that decay via gravitational radiation or other channels such as Goldstone boson emission. These mechanisms introduce energy loss terms into the evolution equations, modifying the scaling behavior and preventing the string network from dominating the energy budget of the Universe. Here, we do not include the possibility for emissions and we assume that the strings dilute according to their cosmological abundance as in eq.~\eqref{eq:freestrings}.

The combined equation of state for a two-component dark fluid consisting of a cosmological constant and a string network is given by
\begin{equation}
    w_{\mathrm{DE}} = \frac{-\Omega_\Lambda + \frac{1}{3}(2v_{\mathrm{s}}^2 - 1)\Omega_{\mathrm{s}}\,a^{-2(1+v_{\mathrm{s}}^2)}}{\Omega_\Lambda + \Omega_{\mathrm{s}}\,a^{-2(1+v_{\mathrm{s}}^2)}}\,,
    \label{eq:dark_energy_eos}
\end{equation}
where $\Omega_\Lambda$ and $\Omega_{\mathrm{s}}$ are the present-day density parameters for the cosmological constant and string network, respectively. These quantities are expressed in units of the critical density today, $\rho_\mathrm{crit} = 3H_0^2/(8\pi G)$. With this parametrization, the energy density components at the present time in a flat Universe must satisfy the constraint
\begin{equation}
    \Omega_\Lambda + \Omega_{\mathrm{m}} + \Omega_{\mathrm{r}} + \Omega_{\mathrm{s}} = 1\,,
\end{equation}
where $\Omega_{\mathrm{m}}$ and $\Omega_{\mathrm{r}}$ are the density parameters for matter and radiation, respectively. In what follows, we analyze the cosmological consequences of this string-inspired dark fluid scenario and derive observational constraints on the parameters $\Omega_{\mathrm{s}}$ and $v_{\mathrm{s}}$.

For the sake of clarity, we derive a general expression for the dark fluid density as a function of the scale factor, using the continuity equation. This is
\begin{equation}
    \rho_{\mathrm{DE}}(a) = \rho_{\mathrm{DE}}(a_0)\, a^{-3} \,\exp\left[-3\int_1^a \frac{w_{\mathrm{DE}}(a')}{a'}\, {\rm d}a'\right]\,,
    \label{eq:de_general}
\end{equation}
where $\rho_{\mathrm{DE}}(a_0)$ is the value of the density at present when the scale factor is $a_0$. For future use, we also introduce the present abundance of the dark fluid $\Omega_{\mathrm{DE},0} \equiv \rho_{\mathrm{DE}}(a_0) / \rho_\mathrm{crit}$. After integration, the energy density of the dark fluid is
\begin{equation}
    \rho_{\mathrm{DE}}(a) = \rho_{\mathrm{DE}}(a_0)\,\frac{\Omega_\Lambda + \Omega_{\mathrm{s}}\, a^{-2(1+v_{\mathrm{s}}^2)}}{\Omega_\Lambda + \Omega_{\mathrm{s}}}\,.
    \label{eq:de_rho}
\end{equation}

\begin{table*}[htbp]
\centering
\begin{tabular}{@{}l@{\hspace{1.5cm}}c@{\hspace{1.5cm}}c@{\hspace{1.5cm}}c@{\hspace{1.5cm}}c@{}}
\toprule
Parameter & Model~1 & Model~2 & Model~3 & Model~4 \\
\midrule
$\Omega_b h^2$ & $[0.005, 0.1]$ & $[0.005, 0.1]$ & $[0.005, 0.1]$ & $[0.005, 0.1]$ \\
$\Omega_c h^2$ & $[0.001, 0.99]$ & $[0.001, 0.99]$ & $[0.001, 0.99]$ & $[0.001, 0.99]$ \\
$\tau_{\mathrm{reio}}$ & $[0.01, 0.8]$ & $[0.01, 0.8]$ & $[0.01, 0.8]$ & $[0.01, 0.8]$ \\
$100\theta_s$ & $[0.5, 10]$ & $[0.5, 10]$ & $[0.5, 10]$ & $[0.5, 10]$ \\
$\ln(10^{10}A_{\mathrm{s}})$ & $[1.61, 3.91]$ & $[1.61, 3.91]$ & $[1.61, 3.91]$ & $[1.61, 3.91]$ \\
$n_{\mathrm{s}}$ & $[0.8, 1.2]$ & $[0.8, 1.2]$ & $[0.8, 1.2]$ & $[0.8, 1.2]$ \\
$\Omega_{\mathrm{s}}$ & $[0, 1]$ & $[0, 1]$ & $[-0.5, 1]$ & $[-0.5, 1]$ \\
$v_{\mathrm{s}}$ & --- & $[0, 0.9]$ & --- & $[0, 0.9]$ \\
\bottomrule
\end{tabular}
\caption{Prior ranges for free cosmological parameters. See the text for the descriptions.}
\label{tab:priors}
\end{table*}

\section{Methodology and Data}
\label{sec:methods}

To perform parameter inference, we employ the \texttt{Cobaya} framework~\cite{Torrado:2020dgo}, which implements a Markov Chain Monte Carlo (MCMC) sampler tailored for cosmological applications. This is used in conjunction with a modified version of \texttt{CAMB}~\cite{Lewis:1999bs}, a Boltzmann solver that we adapt to include the DDE parametrization. Dark energy perturbations are modeled using the default parameterized post-Friedmann (PPF) approach implemented in \texttt{CAMB}~\cite{Lewis:1999bs}. We assess the convergence of the MCMC chains using the Gelman--Rubin statistic $R-1$~\cite{gelman_inference_1992}, adopting a convergence threshold of $R - 1 < 0.01$. The posterior distributions and parameter contours are analyzed and visualized using the \texttt{getdist} package~\cite{Lewis:2019xzd}.

To systematically compare the cosmic string models under consideration, we make use of the following observational datasets:
\begin{itemize}
    \item \textbf{CMB:} Measurements of the Cosmic Microwave Background (CMB) from the \textit{Planck} 2018 legacy release, including the high-$\ell$ Plik TT, TE, and EE likelihoods, the low-$\ell$ TT-only Commander likelihood, and the low-$\ell$ EE-only SimAll likelihood~\cite{Planck:2018nkj,Planck:2019nip,Planck:2018lbu}. We also incorporate lensing data from the Atacama Cosmology Telescope (ACT) DR6, specifically the \texttt{actplanck baseline} likelihood~\cite{ACT:2023kun,ACT:2023dou}. This combined dataset is referred to collectively as \textbf{CMB}.
    
    \item \textbf{BAO:} Baryon Acoustic Oscillation (BAO) and Redshift-Space Distortion (RSD) measurements from the completed SDSS-IV eBOSS survey~\cite{eBOSS:2020yzd}, which includes isotropic and anisotropic distance indicators and expansion rate constraints across a wide redshift range, incorporating Lyman-$\alpha$ BAO measurements. This is referred to as \textbf{SDSS}. We also use BAO data from the first three years of the Dark Energy Spectroscopic Instrument (DESI DR2)~\cite{DESI:2025zgx,DESI:2025fii,DESI:2025qqy}, labeled as \textbf{DESI}.
    
    \item \textbf{Type Ia Supernovae:} Distance modulus measurements from Type Ia Supernovae (SNe Ia) provided by the PantheonPlus sample~\cite{Scolnic:2021amr,Brout:2022vxf}, consisting of 1701 light curves from 1550 unique SNe Ia over the redshift range $z \in [0.001,\, 2.26]$. This dataset is denoted as \textbf{PantheonPlus}. Additionally, we include the complete five-year data release from the Dark Energy Survey (DES), comprising 1635 SNe Ia with redshifts in the range $0.1 < z < 1.13$~\cite{DES:2024hip,DES:2024jxu,DES:2024upw}, referred to as \textbf{DESY5}.
\end{itemize}

In our study, we employ flat and uniform priors as detailed in table~\ref{tab:priors}. The extended cosmic string models are formulated as extensions of the standard six-parameter $\Lambda$CDM framework. This baseline model includes the baryon density $\Omega_b h^2$, cold dark matter density $\Omega_c h^2$, optical depth to reionization $\tau_{\mathrm{reio}}$, the scalar fluctuation amplitude and spectral index $\ln(10^{10} A_{\mathrm{s}})$ and $n_{\mathrm{s}}$, and the angular size of the sound horizon at recombination $\theta_s$. To this, we add the cosmic string energy density $\Omega_{\mathrm{s}}$, which is allowed to take only positive values in prior sets~1 and~2, and both positive and negative values in prior sets~3 and~4. Additionally, the string network's bulk velocity is introduced as a free parameter in prior sets~2 and~4.

Finally, we perform Bayesian model comparison by computing the logarithm of the Bayesian evidence $\ln \mathcal{Z}$ using \texttt{MCEvidence}~\cite{Heavens:2017afc}, accessed via the {\tt Cobaya} wrapper provided in the \texttt{wgcosmo} repository~\cite{giare2025wgcosmo}. For each model $\mathcal{M}_i$ with parameter vector $\Theta$, Bayes' theorem gives the posterior as:
\begin{equation}
    P(\Theta|D, \mathcal{M}_i) = \frac{\mathcal{L}(D|\Theta, \mathcal{M}_i)\, \pi(\Theta|\mathcal{M}_i)}{\mathcal{Z}_i},
    \label{eq:bayes_theorem}
\end{equation}
where $\mathcal{L}$ is the likelihood, $\pi$ the prior, and the evidence $\mathcal{Z}_i$ is given by:
\begin{equation}
    \mathcal{Z}_i = \int \mathcal{L}(D|\Theta, \mathcal{M}_i)\, \pi(\Theta|\mathcal{M}_i)\, {\rm d}\Theta\,.
    \label{eq:bayesian_evidence}
\end{equation}

To compare models, we compute the Bayes factor $\mathcal{Z}_{ij} = \mathcal{Z}_i / \mathcal{Z}_j$ and define the relative log-evidence as:
\begin{equation}
    \Delta \ln \mathcal{Z}_{ij} \equiv \ln \mathcal{Z}_i - \ln \mathcal{Z}_j\,,
    \label{eq:relative_log_bayesian_evidence}
\end{equation}
where $i$ refers to one of the cosmic string models and $j$ to the baseline $\Lambda$CDM case. Positive values of $\Delta \ln \mathcal{Z}_{ij}$ indicate a preference for the extended model.

We interpret $\Delta \ln \mathcal{Z}_{ij}$ using the revised Jeffreys' scale~\cite{Kass:1995loi}: values in the range $[0, 1]$ are \textit{inconclusive}, $[1, 2.5]$ indicate \textit{weak} evidence, $[2.5, 5]$ \textit{moderate}, $[5, 10]$ \textit{strong}, and values above $10$ indicate \textit{very strong} evidence in favor of the preferred model.

%%%%%%%%%%%%%%%%%%%%%%%%%%%%%%%%%%%%%%%%%%%%%%%%%%%%%%%
\begin{table*}[htbp]
\normalsize
\centering
\resizebox{\textwidth}{!}{%
\begin{tabular}{l|ccccccc}
\toprule
\textbf{Parameters} 
& \textbf{CMB} 
& \textbf{CMB} 
& \textbf{CMB} 
& \textbf{CMB+SDSS}
& \textbf{CMB} 
& \textbf{CMB} 
& \textbf{CMB+DESI} \\
&
& \textbf{+SDSS} 
& \textbf{+PantheonPlus} 
& \textbf{+PantheonPlus}
& \textbf{+DESI} 
& \textbf{+DESI+DESY5} 
& \textbf{+PantheonPlus} \\
\midrule
$\Omega_b\,h^2$ 
& $0.02235 \pm 0.00015$ 
& $0.02247 \pm 0.00014$ 
& $0.02235 \pm 0.00014$ 
& $0.02245 \pm 0.00013$
& $0.02257 \pm 0.00013$ 
& $0.02254 \pm 0.00013$ 
& $0.02255 \pm 0.00012$ \\[6pt]

$\Omega_c\,h^2$ 
& $0.1205 \pm 0.0012$ 
& $0.11890 \pm 0.00091$ 
& $0.1203 \pm 0.0011$ 
& $0.11915 \pm 0.00090$
& $0.11751 \pm 0.00069$ 
& $0.11781^{+0.00075}_{-0.00067}$ 
& $0.11767 \pm 0.00068$ \\[6pt]

$100\,\theta_s$ 
& $1.04088 \pm 0.00031$ 
& $1.04109 \pm 0.00029$ 
& $1.04089 \pm 0.00030$ 
& $1.04105 \pm 0.00029$
& $1.04126 \pm 0.00027$ 
& $1.04124 \pm 0.00028$ 
& $1.04125 \pm 0.00028$ \\[6pt]

$\tau_{\mathrm{reio}}$ 
& $0.0557 \pm 0.0075$ 
& $0.0602^{+0.0066}_{-0.0080}$ 
& $0.0554^{+0.0069}_{-0.0077}$ 
& $0.0597^{+0.0069}_{-0.0078}$
& $0.0634^{+0.0071}_{-0.0082}$ 
& $0.0626^{+0.0069}_{-0.0081}$ 
& $0.0629^{+0.0071}_{-0.0079}$ \\[6pt]

$n_{\mathrm{s}}$ 
& $0.9645 \pm 0.0042$ 
& $0.9685 \pm 0.0037$ 
& $0.9647 \pm 0.0040$ 
& $0.9678 \pm 0.0037$
& $0.9719 \pm 0.0034$ 
& $0.9711 \pm 0.0034$ 
& $0.9714 \pm 0.0034$ \\[6pt]

$\ln\bigl(10^{10}\,A_{\mathrm{s}}\bigr)$ 
& $3.050 \pm 0.014$ 
& $3.056^{+0.012}_{-0.014}$ 
& $3.049 \pm 0.014$ 
& $3.055 \pm 0.014$
& $3.061^{+0.013}_{-0.015}$ 
& $3.060^{+0.013}_{-0.015}$ 
& $3.060^{+0.013}_{-0.014}$ \\[6pt]

$\Omega_{\mathrm{s}}$ 
& $< 0.0465$ 
& $< 0.0103$ 
& $< 0.0257$ 
& $< 0.0113$
& $< 0.00824$ 
& $< 0.0114$ 
& $< 0.00942$ \\[6pt]
\midrule

$H_0$ [km/s/Mpc] 
& $65.9^{+1.5}_{-0.72}$ 
& $67.56 \pm 0.46$ 
& $66.35^{+0.83}_{-0.66}$ 
& $67.40 \pm 0.45$
& $68.26 \pm 0.32$ 
& $68.00 \pm 0.33$ 
& $68.13^{+0.32}_{-0.29}$ \\[6pt]

$\sigma_{8}$ 
& $0.800^{+0.014}_{-0.0067}$ 
& $0.8093 \pm 0.0060$ 
& $0.8037^{+0.0092}_{-0.0069}$ 
& $0.8091^{+0.0064}_{-0.0056}$
& $0.8080 \pm 0.0060$ 
& $0.8071^{+0.0066}_{-0.0059}$ 
& $0.8076 \pm 0.0061$ \\[6pt]

$S_{8}$ 
& $0.839 \pm 0.013$ 
& $0.8242 \pm 0.0097$ 
& $0.837 \pm 0.011$ 
& $0.8267 \pm 0.0094$
& $0.8108 \pm 0.0079$ 
& $0.8137 \pm 0.0079$ 
& $0.8122 \pm 0.0078$ \\[6pt]

$\Omega_{\mathrm{m}}$ 
& $0.3306^{+0.0087}_{-0.018}$ 
& $0.3112 \pm 0.0058$ 
& $0.3258^{+0.0081}_{-0.0095}$ 
& $0.3132 \pm 0.0056$
& $0.3020 \pm 0.0039$ 
& $0.3050 \pm 0.0037$ 
& $0.3035 \pm 0.0037$ \\[6pt]
\midrule

$\Delta \chi^2_{\mathrm{min},\,\Lambda \mathrm{CDM}}$ 
& $-0.31$ 
& $1.09$ 
& $-0.64$ 
& $0.65$
& $0.98$ 
& $0.04$ 
& $0.84$ \\[6pt]

$\Delta \ln \mathcal{Z}_{\,\Lambda {\rm CDM}}$ 
& $-3.94$ 
& $-5.4$ 
& $-4.13$ 
& $-5.23$
& $-5.69$ 
& $-5.2$ 
& $-5.53$ \\[6pt]

\bottomrule
\end{tabular}
}
\caption{Observational constraints at 68\% CL and upper limits at 95\% CL obtained for Model~1 using various datasets. Negative values of $\Delta \chi^2_{\mathrm{min},\,\Lambda \mathrm{CDM}}$ favor our model over the standard $\Lambda$CDM scenario, while positive values of $\Delta \ln \mathcal{Z}_{\,\Lambda \mathrm{CDM}}$ indicate a preference for our model.}
\label{tab:model1}
\end{table*}

\begin{figure*}[htbp]
    \centering
    \includegraphics[width=0.9\linewidth]{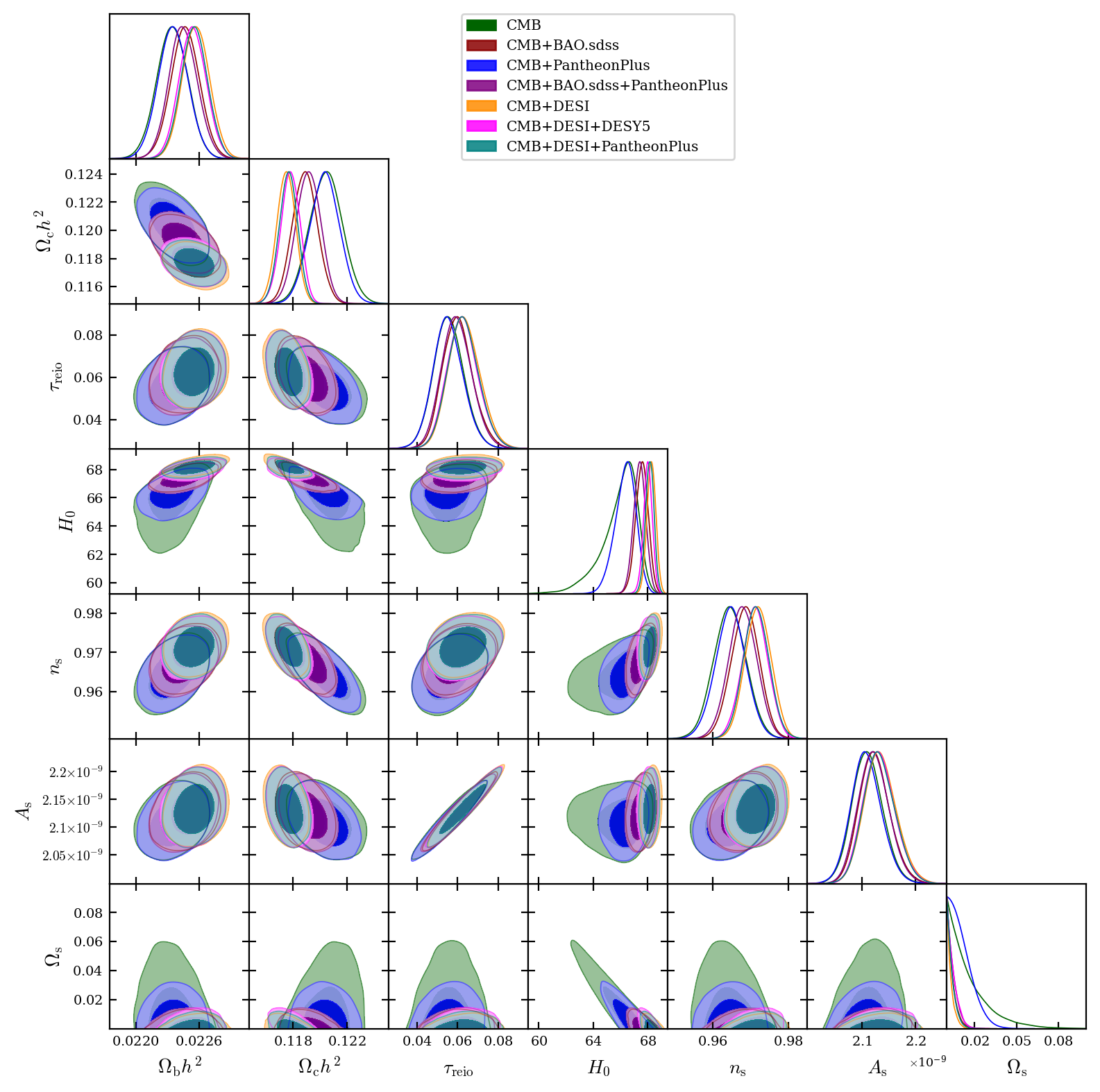}
\caption{One-dimensional posterior distributions and two-dimensional marginalized contours for Model~1, obtained using different combinations of cosmological datasets.}
    \label{fig:Post_P1}
\end{figure*}

\begin{figure*}[t!]
    \centering
    \begin{subfigure}[t]{0.49\textwidth}
        \centering
        \includegraphics[width=0.99\linewidth]{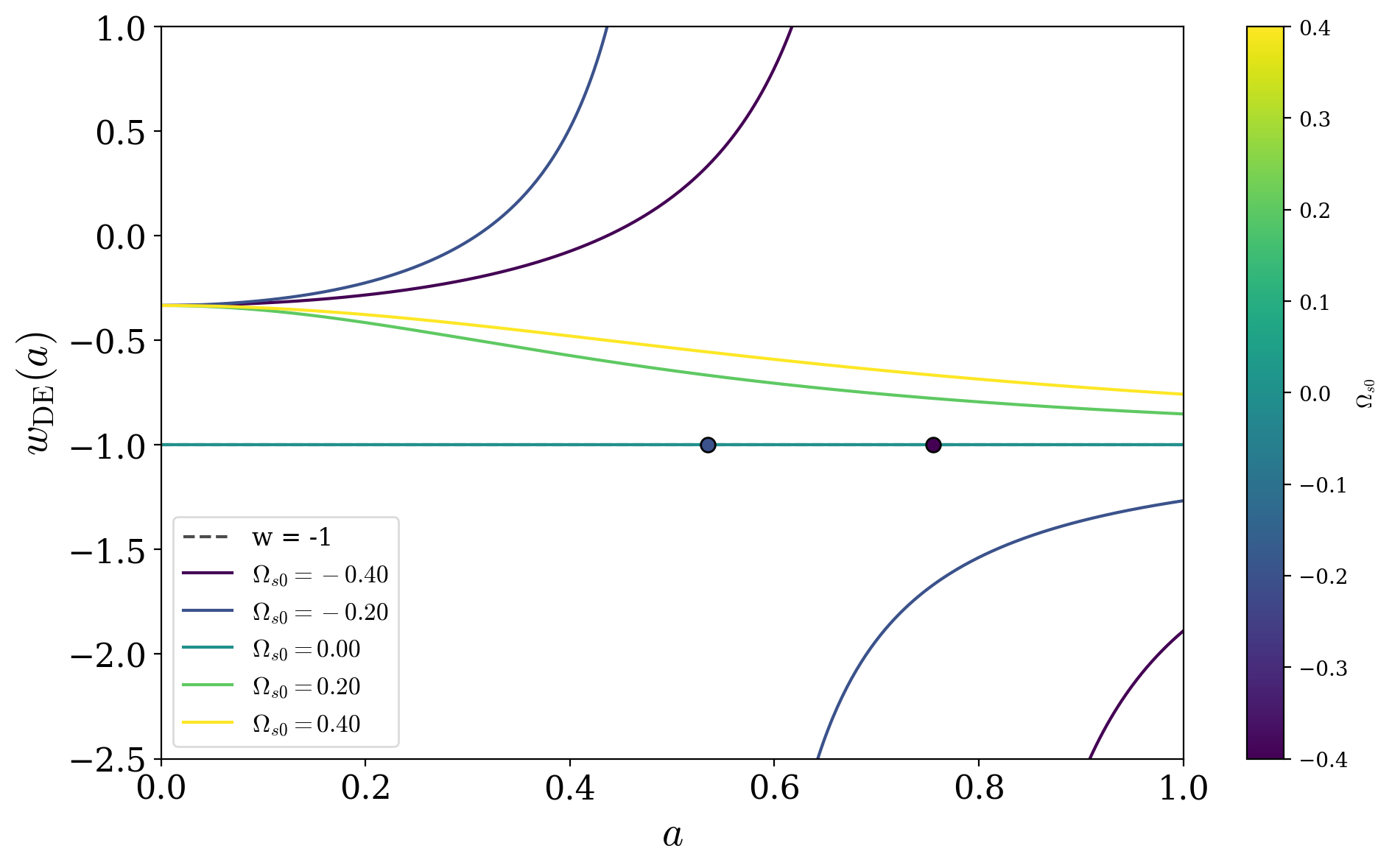}
    \end{subfigure}
    \begin{subfigure}[t]{0.49\textwidth}
        \centering
        \includegraphics[width=0.99\linewidth]{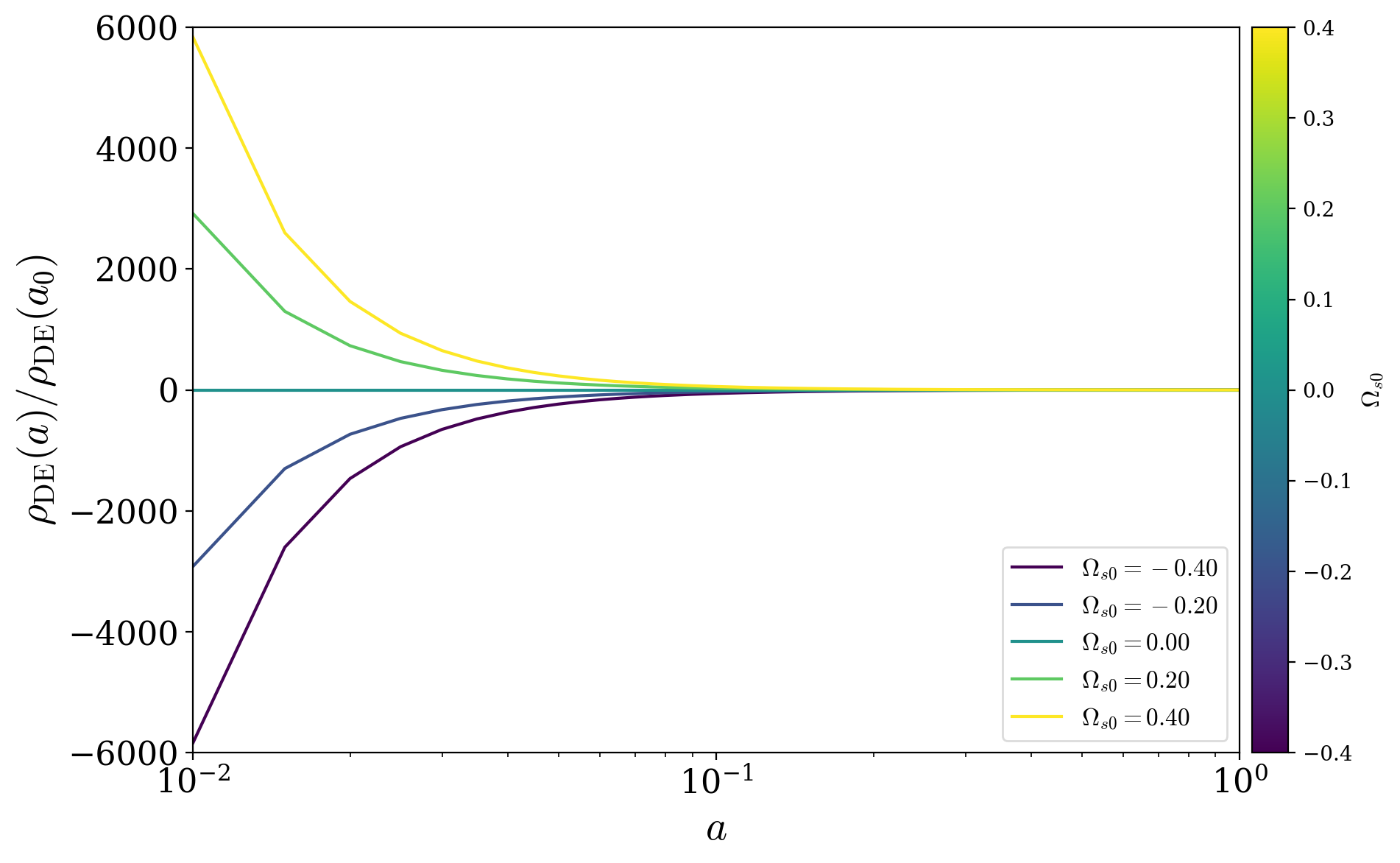}
    \end{subfigure}
    \includegraphics[width=0.5\linewidth]{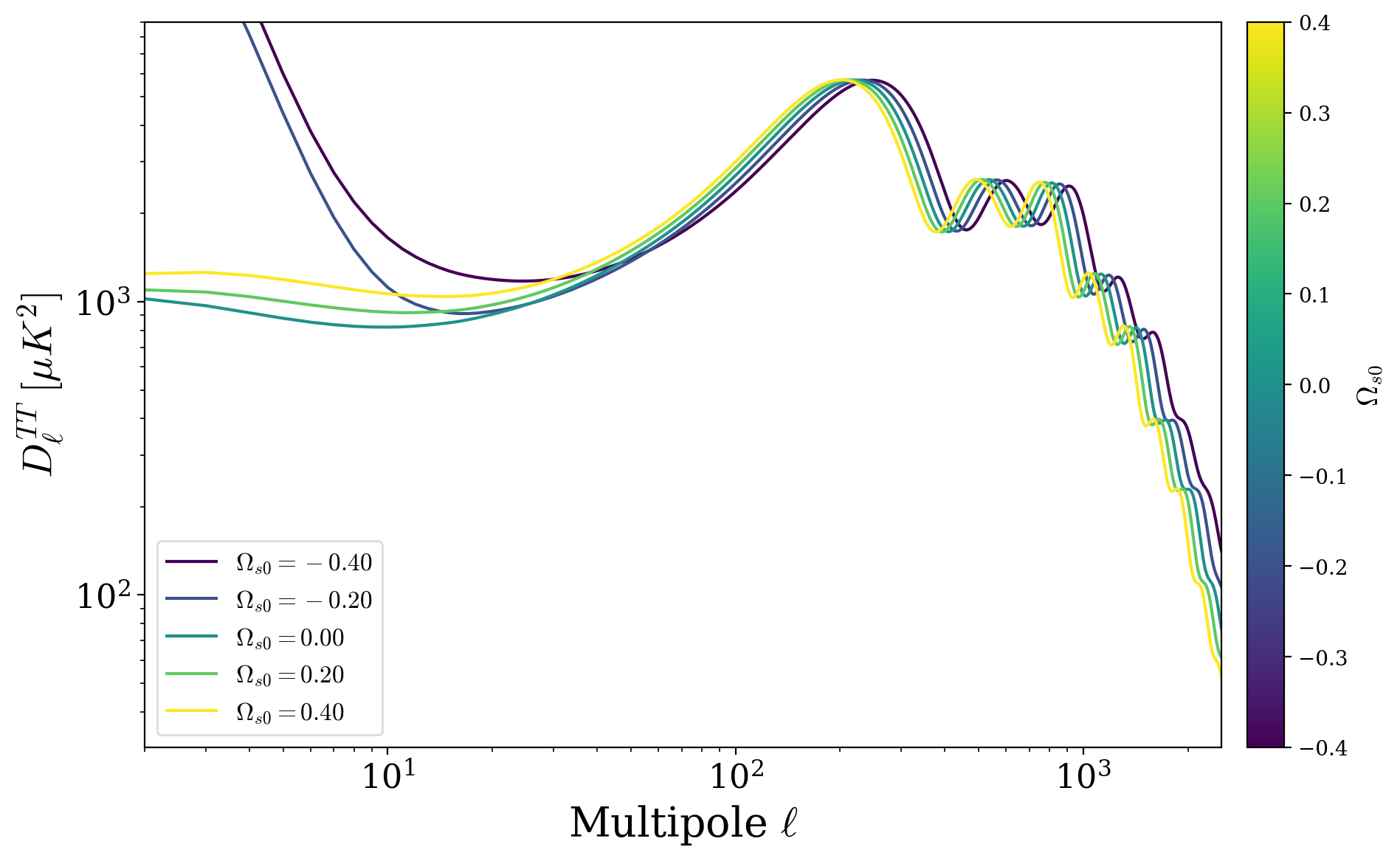}
\caption{{\bf Model~1: Non-relativistic cosmic string network.} \textit{Top panel:} The equation of state $w_{\rm DE}$ from eq.~\eqref{eq:dark_energy_eos} (left) and the energy density ratio $\rho_{\rm DE}(a)/\rho_{\rm DE}(a_0)$ from eq.~\eqref{eq:de_rho} (right), shown as functions of the scale factor $a$. For illustration, the present-day DE contribution is fixed at $\Omega_{\mathrm{DE},0} = 0.7$. \textit{Bottom panel:} The impact of varying the modified DE parameter $\Omega_{\mathrm{s}}$ on the CMB TT power spectrum.}
    \label{fig:w_de_first}
\end{figure*}

\begin{figure*}[htbp]
    \begin{center}
    \includegraphics[width=0.9\linewidth]{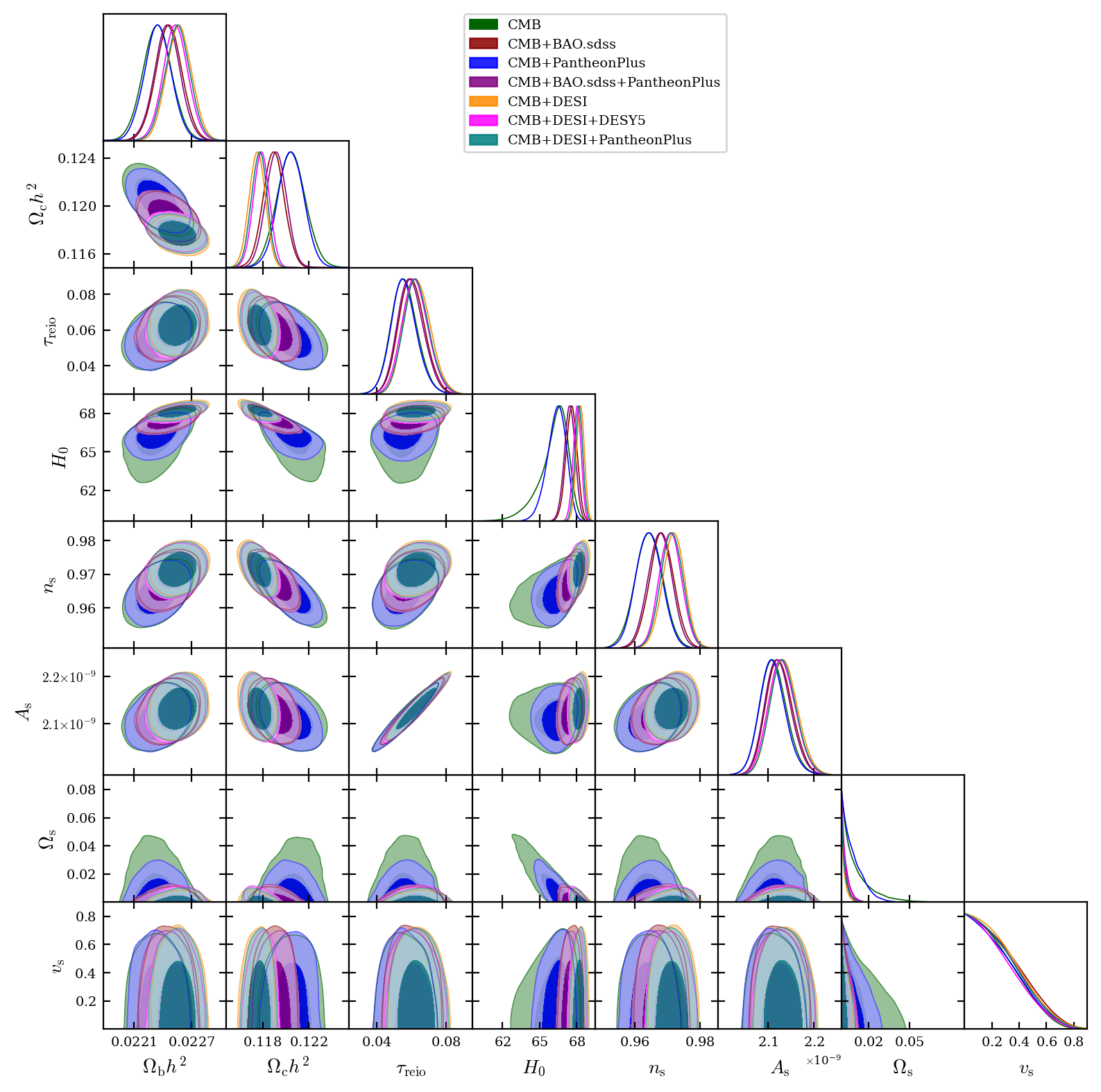}
    \end{center}
\caption{One-dimensional posterior distributions and two-dimensional marginalized contours for Model~2, derived using different combinations of cosmological datasets. The contours illustrate parameter degeneracies and the progressive tightening of constraints as additional data are included.}
    \label{fig:Post_P2}
\end{figure*}

\begin{figure*}[t!]
    \centering
    \begin{subfigure}[t]{0.49\textwidth}
        \centering
        \includegraphics[width=0.99\linewidth]{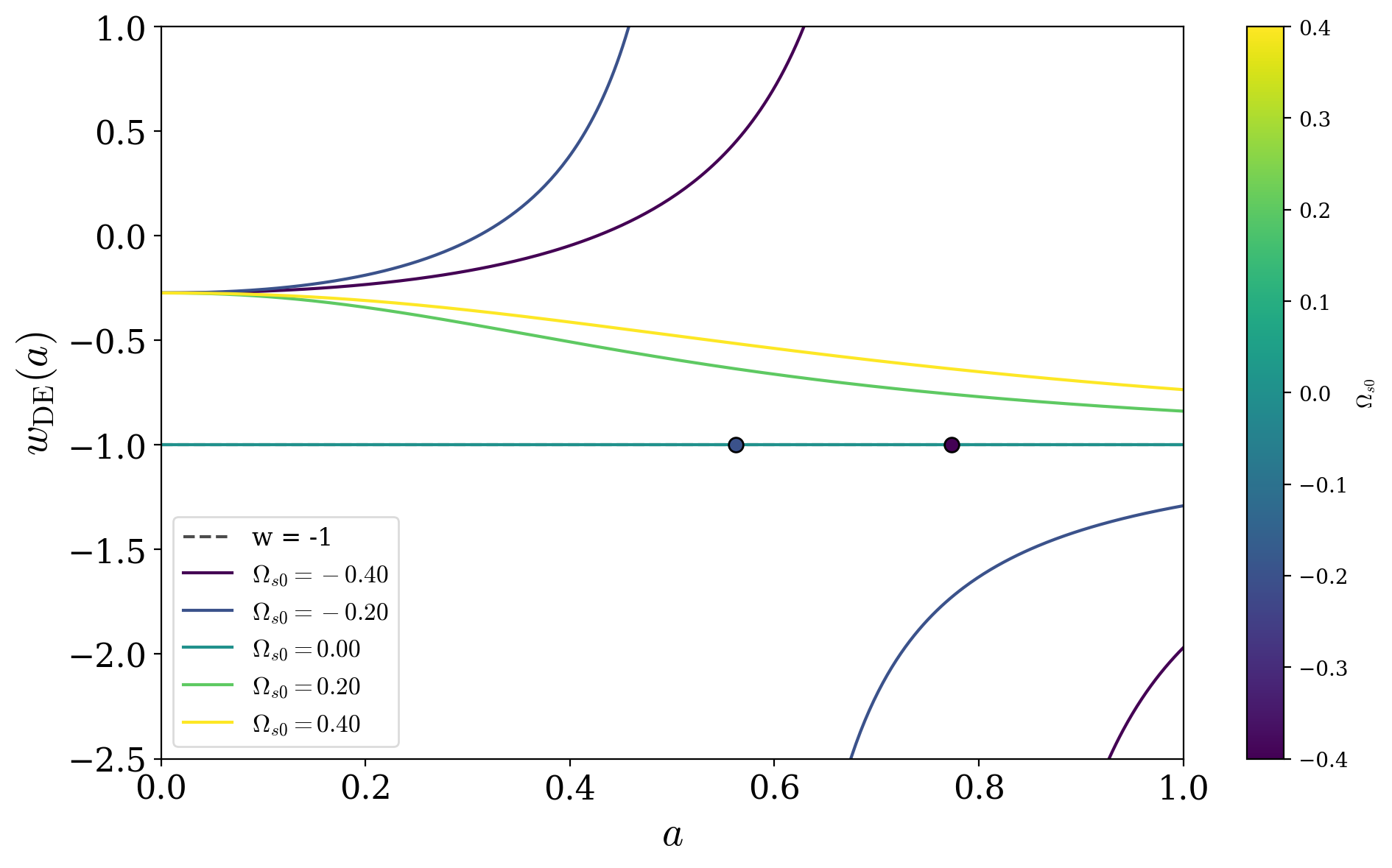}
    \end{subfigure}
    \begin{subfigure}[t]{0.49\textwidth}
        \centering
        \includegraphics[width=0.99\linewidth]{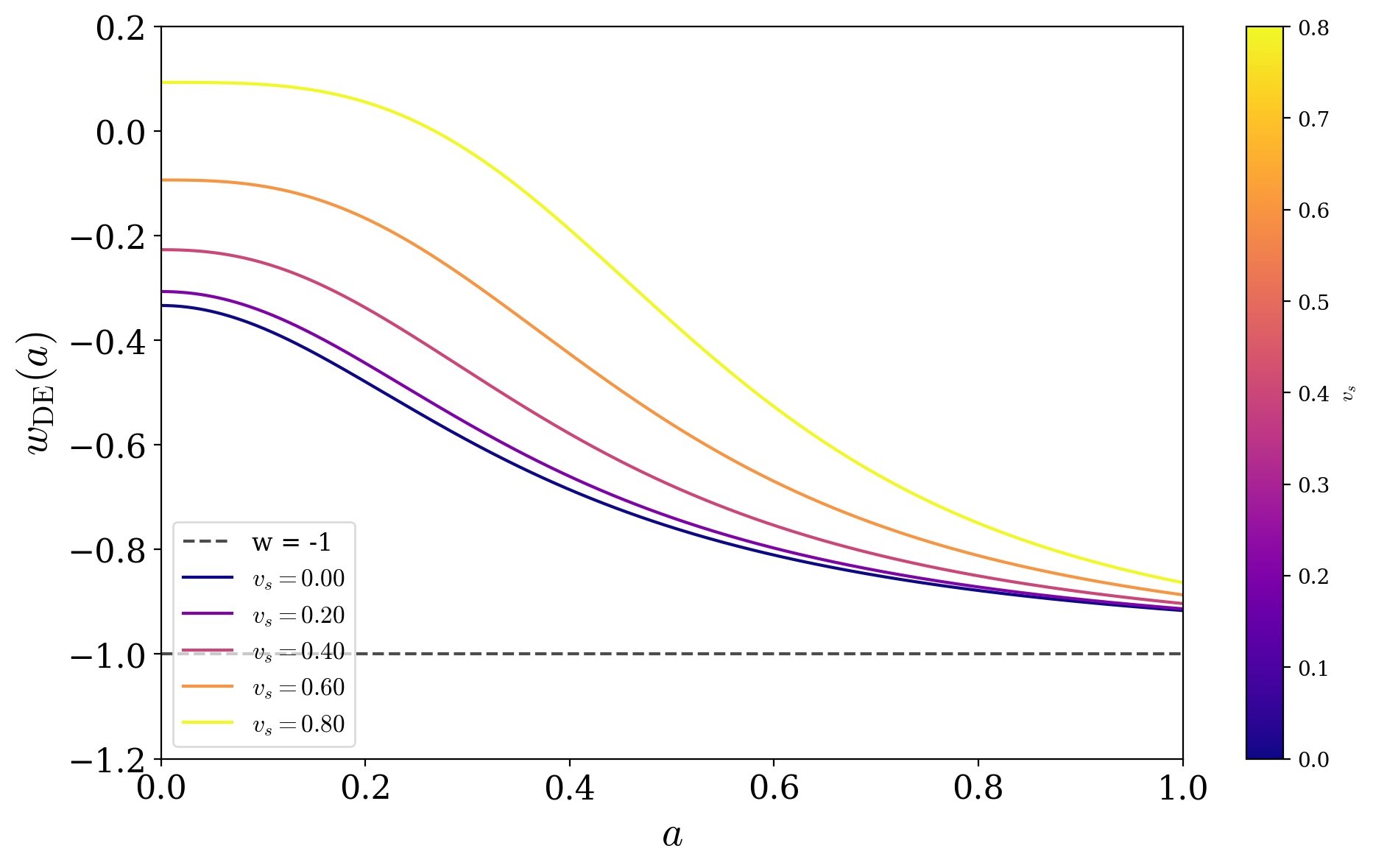}
    \end{subfigure}
    \begin{subfigure}[t]{0.49\textwidth}
        \centering
        \includegraphics[width=0.99\linewidth]{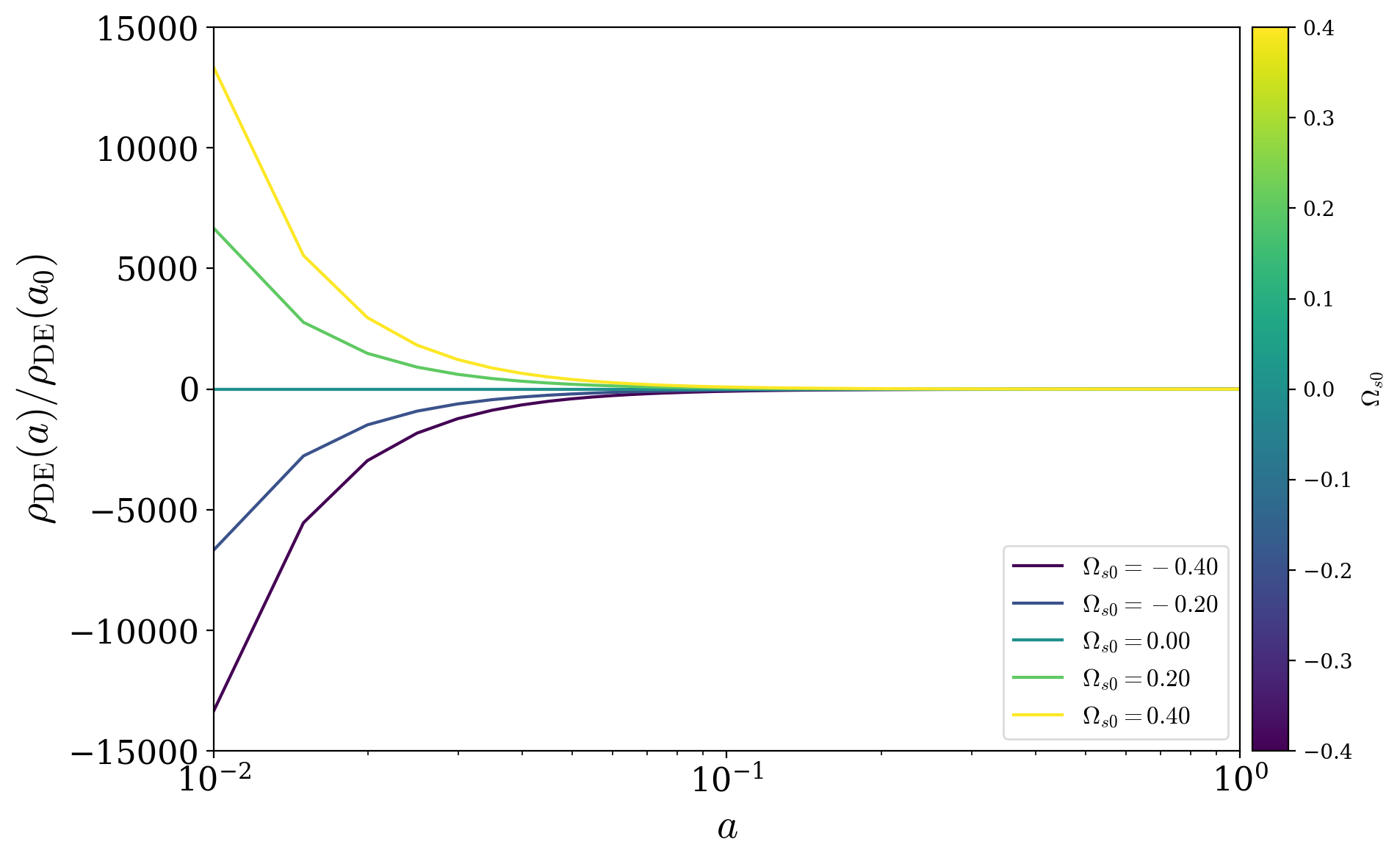}
    \end{subfigure}
    \begin{subfigure}[t]{0.49\textwidth}
        \centering
        \includegraphics[width=0.99\linewidth]{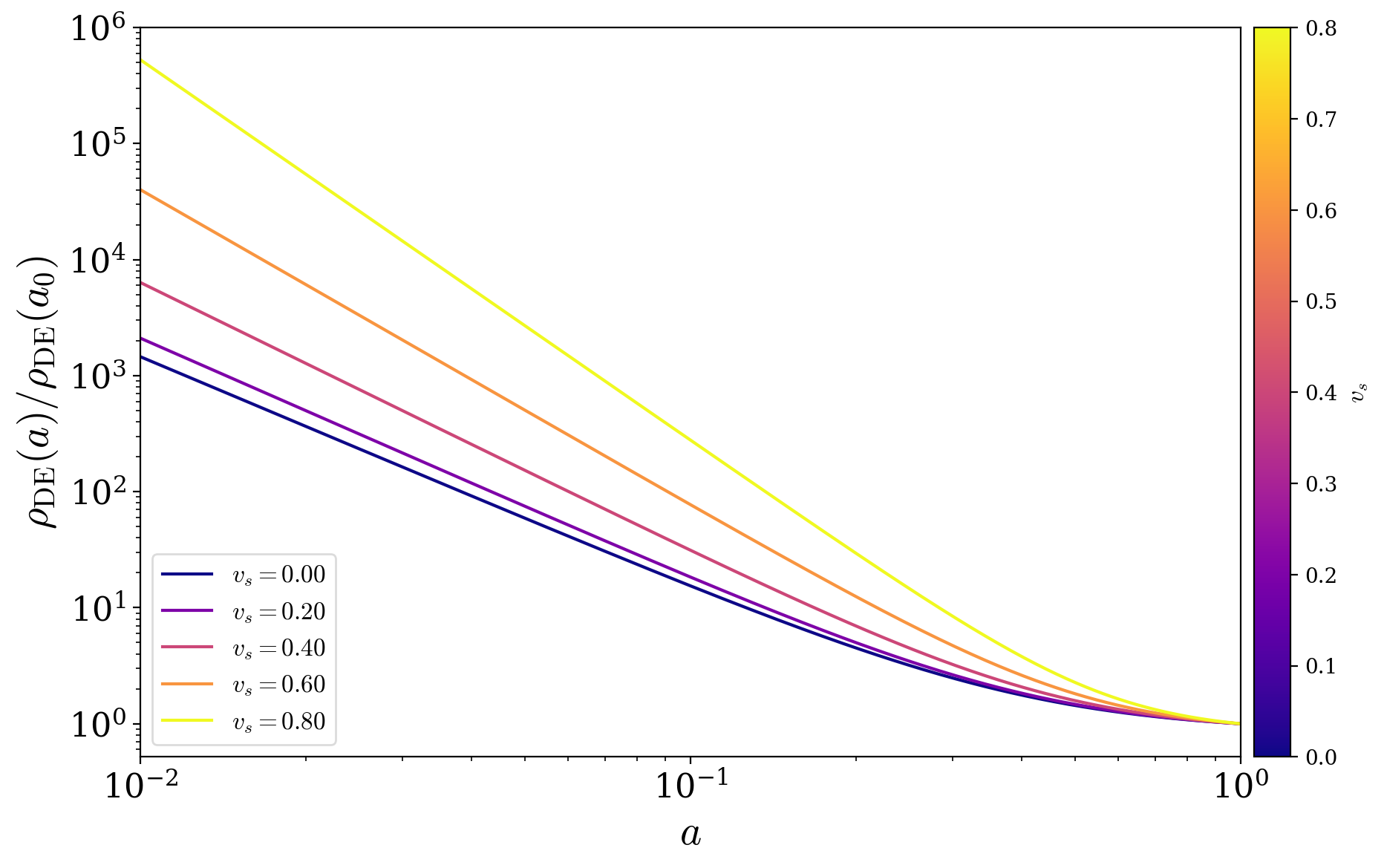}
    \end{subfigure}
    \begin{subfigure}[t]{0.49\textwidth}
        \centering
        \includegraphics[width=0.99\linewidth]{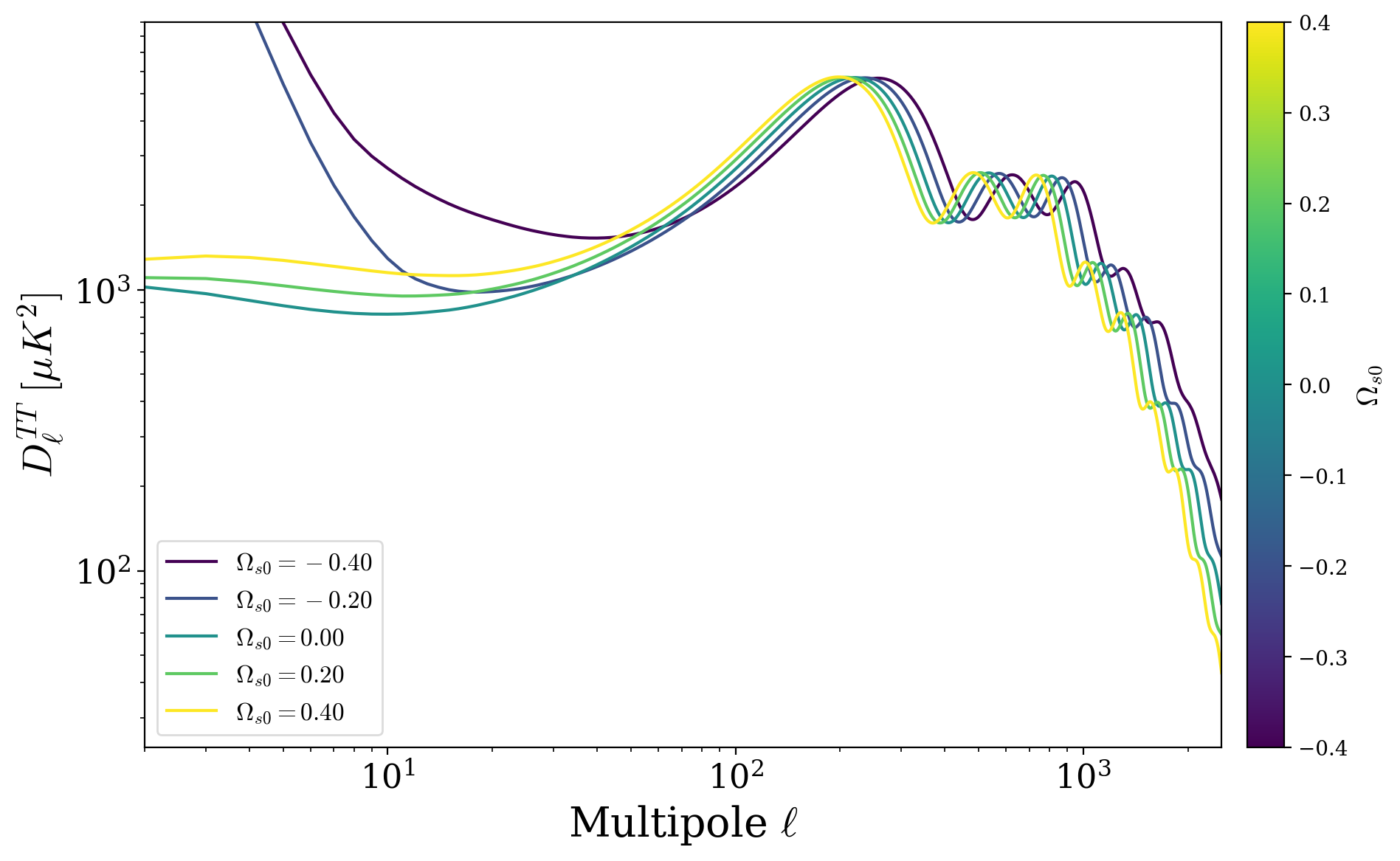}
    \end{subfigure}
    \begin{subfigure}[t]{0.49\textwidth}
        \centering
        \includegraphics[width=0.99\linewidth]{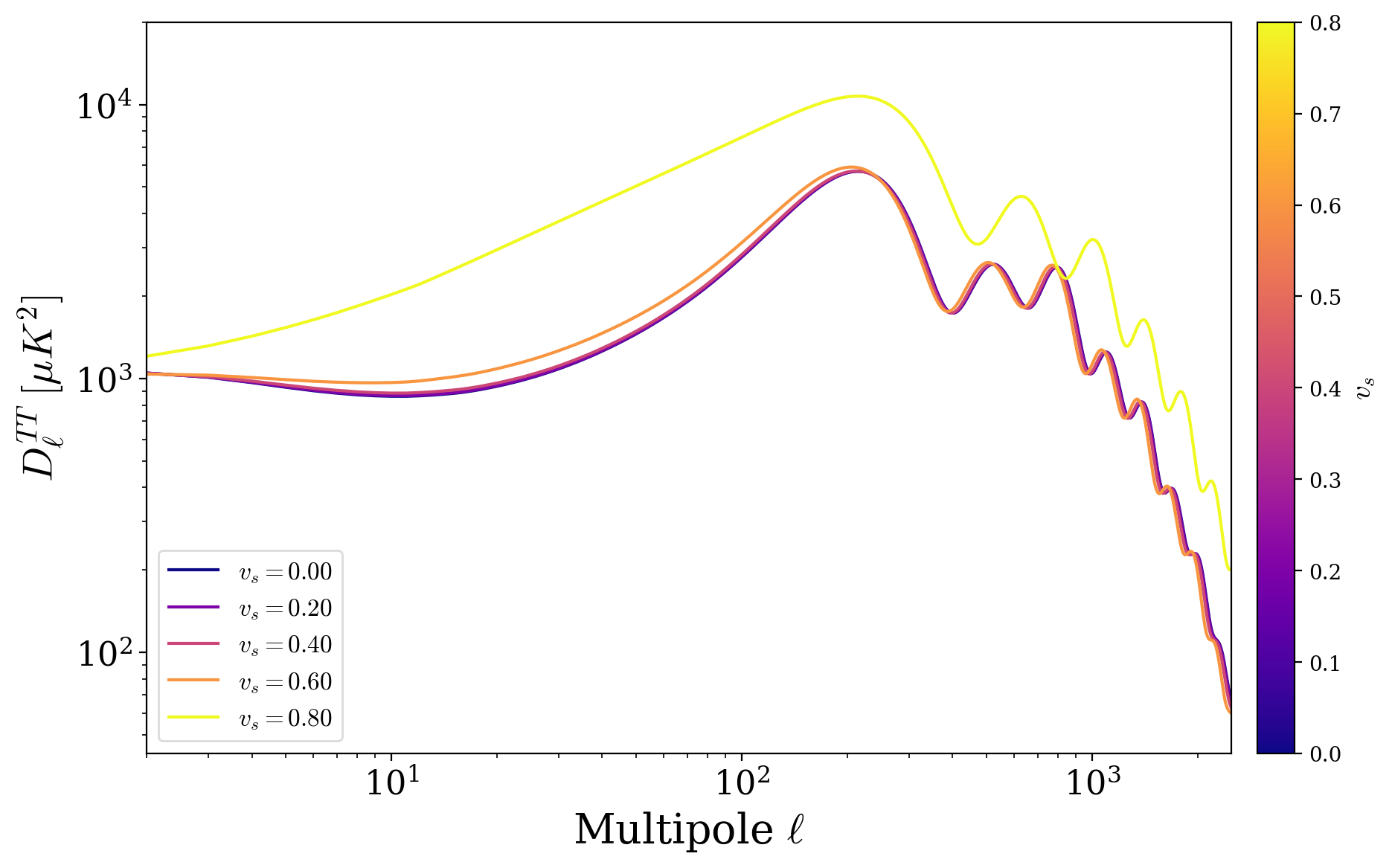}
    \end{subfigure}
\caption{{\bf Model~2: Velocity-dependent string network.} \textit{Top panel:} The equation of state $w_{\rm DE}$ from eq.~\eqref{eq:dark_energy_eos} as a function of the scale factor $a$. \textit{Middle panel:} The energy density ratio $\rho_{\rm DE}(a)/\rho_{\rm DE}(a_0)$ from eq.~\eqref{eq:de_rho}, also shown as a function of $a$. For both the top and middle panels, the left side shows fixed $v_{\mathrm{s}} = 0.3$ with varying $\Omega_{\mathrm{s}}$ as indicated in the legend, while the right side shows fixed $\Omega_{\mathrm{s}} = 0.1$ with varying $v_{\mathrm{s}}$. The present-day DE contribution is fixed at $\Omega_{\mathrm{DE},0} = 0.7$ for illustration. \textit{Bottom panel:} The impact on the CMB TT power spectrum when varying $\Omega_{\mathrm{s}}$ with fixed $v_{\mathrm{s}} = 0.3$ (left), and when varying $v_{\mathrm{s}}$ with fixed $\Omega_{\mathrm{s}} = 0.1$ (right), where $\Omega_{\mathrm{DE},0}$ is defined as $\Omega_{\mathrm{DE},0} = 1 - \Omega_{\mathrm{m}} - \Omega_k - \Omega_{\mathrm{r}}$.}
    \label{fig:w_de_second}
\end{figure*}

\section{Results}\label{results}

\subsection{Constraints on Model~1: Positive energy, non-relativistic string network.}

We first discuss the results for a non-relativistic string network with positive energy density, Model~1, which extends the baseline $\Lambda$CDM framework by including a positive contribution from non-relativistic cosmic strings, parametrized by the energy density $\Omega_{\mathrm{s}}$. The results for this model, derived from various combinations of datasets, are summarized in table~\ref{tab:model1}. The posterior distributions place increasingly stringent upper bounds on $\Omega_{\mathrm{s}}$ as additional data are incorporated. Using CMB data alone, we find $\Omega_{\mathrm{s}} < 0.0465$ at 95\% confidence level (CL), while the combination of CMB with DESI reduces this bound to $\Omega_{\mathrm{s}} < 0.00824$. These constraints demonstrate the power of large-scale structure data in limiting non-standard sources of anisotropies and density perturbations, particularly those that enhance small-scale clustering or modify the background expansion history.

The inclusion of a string component does not significantly shift the posterior distributions of the standard cosmological parameters. In fact, the constraints from CMB-only data are fully consistent with those of the $\Lambda$CDM model, owing to the data's preference for a null value of $\Omega_{\mathrm{s}}$. However, a strong negative correlation between $\Omega_{\mathrm{s}}$ and $H_0$ is evident in figure~\ref{fig:Post_P1}, leading to a lower bound on the Hubble constant from CMB-only data: $H_0 = 65.9^{+1.5}_{-0.72} \,\mathrm{km\,s^{-1}\,Mpc^{-1}}$ at 68\% CL. As a consequence, the matter density shifts toward higher values, with $\Omega_{\mathrm{m}} = 0.3306^{+0.0087}_{-0.018}$ at 68\% CL, due to its strong anticorrelation with $H_0$.

The addition of SDSS data not only strengthens the constraints on $\Omega_{\mathrm{s}}$ by nearly a factor of four but also increases the Hubble constant and decreases the matter density, bringing them back into agreement with a Planck $\Lambda$CDM cosmology. While the inclusion of PantheonPlus alongside CMB has similar effects, they are milder; moreover, adding PantheonPlus to the CMB+SDSS combination yields negligible impact, with constraints essentially unchanged from the CMB+SDSS case.

Notably, the scalar spectral index $n_{\mathrm{s}}$ increases slightly with the addition of DESI data, reaching $n_{\mathrm{s}} = 0.9719 \pm 0.0034$, consistent with the mild degeneracy between $\Omega_{\mathrm{s}}$ and the tilt of the primordial spectrum. The Hubble constant $H_0$ also shifts to higher values across dataset combinations, with CMB+DESI yielding $H_0 = 68.26 \pm 0.32 \,\mathrm{km\,s^{-1}\,Mpc^{-1}}$. This trend moves in the direction of alleviating the tension between Planck and local distance ladder measurements~\cite{Riess:2021jrx, Murakami:2023xuy, Breuval:2024lsv}, although the shift remains insufficient to fully resolve the discrepancy.

The derived matter clustering amplitude, $S_8 \equiv \sigma_8 \sqrt{\Omega_{\mathrm{m}} / 0.3}$, is similarly affected. We observe a downward trend from $S_8 = 0.839 \pm 0.013$ (CMB-only) to $S_8 = 0.8108 \pm 0.0079$ (CMB+DESI), indicating improved agreement with weak lensing measurements, which typically favor lower values of $S_8$ than those predicted by Planck under $\Lambda$CDM. While the addition of PantheonPlus does not affect the results in this case either, the inclusion of DESY5 leads to a slightly lower Hubble constant and a relaxed upper bound on $\Omega_{\mathrm{s}}$, suggesting a possible tension among these datasets within this scenario, likely due to DESY5's preference for a higher matter density.

Despite these improvements, the overall statistical preference for Model~1 is not compelling. The difference in minimum $\chi^2$ relative to $\Lambda$CDM, $\Delta \chi^2_{\mathrm{min},\,\Lambda\mathrm{CDM}}$, is small and compatible with one additional degree of freedom, indicating that the extended model does not provide a better fit to the data. Moreover, the difference in Bayesian evidence, $\Delta \ln \mathcal{Z}_{\,\Lambda \mathrm{CDM}}$, is negative in all cases, with values ranging from $-3.94$ to $-5.69$. These results correspond to moderate to strong Bayesian preference for the standard $\Lambda$CDM model, suggesting that the data do not justify the inclusion of an additional degree of freedom such as $\Omega_{\mathrm{s}}$.
Thus, while Model~1 remains consistent with current observations and yields marginal improvements in derived parameters relevant to known cosmological tensions, the statistical evidence does not support the presence of a positive-energy cosmic string component in the cosmological model.

The features of the cosmic string contribution in Model~1 are illustrated in figure~\ref{fig:w_de_first}. The top panels show the evolution of the dark energy equation of state, $w_{\rm DE}(a)$, and the normalized energy density, $\rho_{\rm DE}(a)/\rho_{\rm DE}(a_0)$, assuming a fixed present-day value of $\Omega_{\mathrm{DE},0} = 0.7$. For positive values of $\Omega_{\mathrm{s}}$, the effective equation of state deviates from that of a cosmological constant ($w = -1$), becoming less negative at early times and asymptotically approaching $w = -1/3$ as $a \to 0$. This behavior reflects the scaling of the string energy density with the scale factor and its role as an early dark energy component, whose relative importance diminishes over time.
The bottom panel quantifies the observable imprint of this modification on the CMB temperature power spectrum. Non-zero $\Omega_{\mathrm{s}}$ enhances power at large angular scales (low multipoles), primarily due to the Integrated Sachs–Wolfe (ISW) effect, while also modifying the acoustic peak structure. These changes are tightly constrained by precision CMB measurements, which explains the strong bounds on $\Omega_{\mathrm{s}}$ reported in table~\ref{tab:model1}. The geometrical degeneracies between $\Omega_{\mathrm{s}}$, $n_{\mathrm{s}}$, and $H_0$, discussed above, manifest as shifts in the peak positions and amplitudes, highlighting the importance of combining CMB with low-redshift data to isolate string-induced deviations.
This behavior is further illustrated in figure~\ref{fig:Post_P1}, which shows one-dimensional posterior distributions and two-dimensional marginalized contours for Model~1 using various combinations of cosmological datasets. The addition of complementary data progressively breaks parameter degeneracies and tightens the constraints on $\Omega_{\mathrm{s}}$, while also inducing shifts in $H_0$, $n_{\mathrm{s}}$, and $\Omega_c h^2$.

\subsection{Constraints on Model~2: Positive energy, velocity-dependent string network.}

Model~2 extends the baseline cosmological scenario by introducing a cosmic string component with positive energy density and allowing the root-mean squared velocity of the string network, $v_{\mathrm{s}}$, to vary as a free parameter. This generalization captures the dynamics of more realistic string networks and enables a test of potential deviations from the assumption of non-relativistic strings. The constraints in table~\ref{tab:model2} show that current cosmological data tightly constrain both the energy density and the velocity of the string network. Using \textit{Planck} CMB data alone, we obtain the upper bounds $\Omega_{\mathrm{s}} < 0.0363$ and $v_{\mathrm{s}} < 0.557$ at 95\% CL. These bounds improve significantly when large-scale structure and supernova data are included. The most stringent constraints come from the \textit{Planck}+DESI combination, yielding $\Omega_{\mathrm{s}} < 0.00663$ and $v_{\mathrm{s}} < 0.592$ at 95\% CL. Interestingly, treating $v_{\mathrm{s}}$ as a free parameter does not weaken the bounds on $\Omega_{\mathrm{s}}$. On the contrary, it strengthens them due to the strong anticorrelation between the two parameters, as illustrated in figure~\ref{fig:Post_P2}. This indicates that the datasets efficiently constrain both parameters simultaneously. It is also noteworthy that the constraints on $v_{\mathrm{s}}$ remain nearly unchanged regardless of the dataset combination used, implying that low-redshift data have little to no effect on this parameter.

%%%%%%%%%%%%%%%%%%%%%%%%%%%%%%%%%%%%%%%%%%%%%%%%%%%%%%%
\begin{table*}[htbp]
\scriptsize
\centering
\resizebox{\textwidth}{!}{%
\begin{tabular}{l|ccccccc}
\toprule
\textbf{Parameters} 
& \textbf{CMB} 
& \textbf{CMB} 
& \textbf{CMB} 
& \textbf{CMB+SDSS}
& \textbf{CMB} 
& \textbf{CMB}
& \textbf{CMB+DESI} \\
&
& \textbf{+SDSS} 
& \textbf{+PantheonPlus} 
& \textbf{+PantheonPlus}
& \textbf{+DESI} 
& \textbf{+DESI+DESY5}
& \textbf{+PantheonPlus} \\
\midrule
$\Omega_b\,h^2$ 
& $0.02235 \pm 0.00015$ 
& $0.02247 \pm 0.00013$ 
& $0.02235 \pm 0.00014$ 
& $0.02245 \pm 0.00013$
& $0.02257 \pm 0.00013$ 
& $0.02253 \pm 0.00013$
& $0.02255 \pm 0.00013$ \\[6pt]

$\Omega_c\,h^2$ 
& $0.1205 \pm 0.0012$ 
& $0.11895 \pm 0.00091$ 
& $0.1204 \pm 0.0011$ 
& $0.11916 \pm 0.00089$
& $0.11752 \pm 0.00069$ 
& $0.11788 \pm 0.00069$
& $0.11770 \pm 0.00068$ \\[6pt]

$100\,\theta_s$ 
& $1.04088 \pm 0.00031$ 
& $1.04108 \pm 0.00029$ 
& $1.04088 \pm 0.00030$ 
& $1.04105 \pm 0.00029$
& $1.04126 \pm 0.00028$ 
& $1.04123 \pm 0.00028$
& $1.04125 \pm 0.00028$ \\[6pt]

$\tau_{\mathrm{reio}}$ 
& $0.0559^{+0.0069}_{-0.0080}$ 
& $0.0603^{+0.0070}_{-0.0079}$ 
& $0.0557^{+0.0068}_{-0.0077}$ 
& $0.0596^{+0.0068}_{-0.0080}$
& $0.0634^{+0.0069}_{-0.0083}$ 
& $0.0622^{+0.0071}_{-0.0080}$
& $0.0629^{+0.0070}_{-0.0079}$ \\[6pt]

$n_{\mathrm{s}}$ 
& $0.9644 \pm 0.0042$ 
& $0.9683 \pm 0.0037$ 
& $0.9644 \pm 0.0040$ 
& $0.9678 \pm 0.0037$
& $0.9720 \pm 0.0034$ 
& $0.9710 \pm 0.0034$
& $0.9714 \pm 0.0034$ \\[6pt]

$\ln\bigl(10^{10}\,A_{\mathrm{s}}\bigr)$ 
& $3.050^{+0.013}_{-0.014}$ 
& $3.056^{+0.013}_{-0.014}$ 
& $3.049^{+0.012}_{-0.014}$ 
& $3.055^{+0.012}_{-0.014}$
& $3.061^{+0.013}_{-0.015}$ 
& $3.059 \pm 0.014$
& $3.060 \pm 0.014$ \\[6pt]

$\Omega_{\mathrm{s}}$ 
& $< 0.0363$ 
& $< 0.00854$ 
& $< 0.0232$ 
& $< 0.00961$
& $< 0.00663$ 
& $< 0.00901$
& $< 0.00779$ \\[6pt]

$v_{\mathrm{s}}$ 
& $< 0.557$ 
& $< 0.604$ 
& $< 0.570$ 
& $< 0.577$
& $< 0.592$ 
& $< 0.569$
& $< 0.588$ \\[6pt]
\midrule

$H_0$ [km/s/Mpc] 
& $66.1^{+1.4}_{-0.66}$ 
& $67.57 \pm 0.45$ 
& $66.35^{+0.86}_{-0.64}$ 
& $67.43 \pm 0.44$
& $68.28 \pm 0.32$ 
& $68.03 \pm 0.32$
& $68.16 \pm 0.32$ \\[6pt]

$\sigma_{8}$ 
& $0.801^{+0.013}_{-0.0065}$ 
& $0.8095 \pm 0.0060$ 
& $0.8038^{+0.0096}_{-0.0066}$ 
& $0.8091 \pm 0.0061$
& $0.8082 \pm 0.0060$ 
& $0.8074 \pm 0.0062$
& $0.8079 \pm 0.0061$ \\[6pt]

$S_{8}$ 
& $0.838 \pm 0.013$ 
& $0.8244 \pm 0.0096$ 
& $0.838 \pm 0.011$ 
& $0.8264 \pm 0.0095$
& $0.8107 \pm 0.0080$ 
& $0.8139 \pm 0.0079$
& $0.8123 \pm 0.0079$ \\[6pt]

$\Omega_{\mathrm{m}}$ 
& $0.3287^{+0.0083}_{-0.016}$ 
& $0.3112 \pm 0.0057$ 
& $0.3260^{+0.0081}_{-0.0098}$ 
& $0.3130 \pm 0.0055$
& $0.3019 \pm 0.0038$ 
& $0.3048 \pm 0.0038$
& $0.3033 \pm 0.0037$ \\[6pt]
\midrule

$\Delta \chi^2_{\mathrm{min},\,\Lambda \mathrm{CDM}}$ 
& $-0.53$ 
& $0.84$ 
& $0.4$ 
& $0.54$
& $1.29$ 
& $-0.1$
& $1.33$ \\[6pt]

$\Delta \ln \mathcal{Z}_{\,\Lambda{\rm CDM}}$ 
& $-4.93$ 
& $-6.18$ 
& $-5.05$ 
& $-6.22$
& $-6.64$ 
& $-6.3$
& $-6.2$ \\[6pt]

\bottomrule
\end{tabular}
}
\caption{Observational constraints at 68\% CL and upper limits at 95\% CL obtained for Model~2 using various datasets. Negative values of $\Delta \chi^2_{\mathrm{min},\,\Lambda \mathrm{CDM}}$ favor Model~2 over the standard $\Lambda$CDM scenario, while positive values of $\Delta \ln \mathcal{Z}_{\,\Lambda \mathrm{CDM}}$ indicate a preference for our model.}
\label{tab:model2}
\end{table*}

The impact on standard cosmological parameters is similar to that observed for Model~1. Due to the strong anticorrelation between $\Omega_{\mathrm{s}}$ and $H_0$, the CMB-only analysis yields a lower value of the Hubble constant and, consequently, a higher matter density. However, the Planck $\Lambda$CDM constraints are recovered once SDSS or PantheonPlus are included in the dataset combination.
The scalar spectral index increases slightly with the inclusion of low-redshift data such as DESI, reaching $n_{\mathrm{s}} = 0.9720 \pm 0.0034$ for \textit{Planck}+DESI. The Hubble parameter also shifts upward, reaching $H_0 = 68.28 \pm 0.32 \,\mathrm{km\,s^{-1}\,Mpc^{-1}}$, again hinting at a mild alleviation of the Hubble tension. Similarly, the clustering amplitude $S_8$ is suppressed to $0.8107 \pm 0.0080$, in better agreement with weak lensing surveys. Despite these phenomenological advantages, statistical model comparison remains unfavorable to Model~2. While the changes in the minimum $\chi^2$ relative to $\Lambda$CDM are small and compatible with two additional degree of freedom, the Bayesian evidence consistently disfavors this extended model. All dataset combinations yield $\Delta \ln \mathcal{Z}_{\,\Lambda \mathrm{CDM}} < -5$, with the strongest result obtained for \textit{Planck}+DESI. These values indicate a strong Bayesian preference for the standard $\Lambda$CDM model over the velocity-dependent cosmic string scenario.

The bounds derived for the root-mean squared string velocity can be compared to theoretical expectations from recent high-resolution simulations of Nambu–Goto and Abelian–Higgs string networks, which predict velocities in the range $v_{\mathrm{s}} \sim 0.3$–$0.6$ during both the radiation- and matter-dominated eras~\cite{Martins:2000cs, Blanco-Pillado:2011egf, Auclair:2019wcv}. This results in a scaling of the energy density slightly faster than $\rho_s \propto a^{-2}$, helping ensure the subdominance of the string network at late times. The upper limits obtained in our analysis, with $v_{\mathrm{s}} \lesssim 0.6$ at 95\% CL across different dataset combinations, are consistent with this predicted range. This agreement suggests that cosmological observations are compatible with a scaling string network, provided its energy density remains sufficiently small to evade current constraints.

To better understand the phenomenology of Model~2, we illustrate in figure~\ref{fig:w_de_second} the behavior of the effective DE component associated with the cosmic string network. The top and middle panels show, respectively, the effective equation of state $w_{\rm DE}(a)$ and the normalized energy density $\rho_{\rm DE}(a)/\rho_{\rm DE}(a_0)$ as functions of the scale factor $a$, derived from Eqs.~\eqref{eq:dark_energy_eos} and~\eqref{eq:de_rho}. For illustrative purposes, we fix the present-day dark energy contribution to $\Omega_{\mathrm{DE},0} = 0.7$.
In the top-left and middle-left panels, we fix the root-mean squared string velocity to $v_{\mathrm{s}} = 0.3$ and vary the present-day string contribution $\Omega_{\mathrm{s}}$, which quantifies the fraction of the effective dark energy density sourced by the string network. As $\Omega_{\mathrm{s}}$ increases, the equation of state $w_{\rm DE}(a)$ becomes less negative at early times due to the increasing contribution from the string component, $\rho_s(a) \propto a^{-2(1+v_{\mathrm{s}}^2)}$. This causes the total DE density to decay more rapidly than that of a cosmological constant and approach the scaling behavior of strings, resulting in $w_{\rm DE}(a) \to \frac{1}{3}(2v_{\mathrm{s}}^2 - 1)$ as $a \to 0$. The corresponding energy density ratio $\rho_{\rm DE}(a)/\rho_{\rm DE}(a_0)$ exhibits stronger evolution with $a$ for higher absolute values of $\Omega_{\mathrm{s}}$, deviating significantly from the cosmological constant solution.
The top-right and middle-right panels fix $\Omega_{\mathrm{s}} = 0.1$ while varying $v_{\mathrm{s}}$. Increasing the string velocity enhances the dynamical impact of the network by making the effective equation of state less negative at earlier times. Since relativistic strings behave more like radiation, this leads to an earlier departure from $w = -1$ and causes a steeper decay of $\rho_{\rm DE}(a)$ with redshift.
The bottom panel of figure~\ref{fig:w_de_second} shows the resulting effect on the CMB temperature power spectrum. In the left panel, we fix $v_{\mathrm{s}} = 0.3$ and vary $\Omega_{\mathrm{s}}$; in the right panel, we fix $\Omega_{\mathrm{s}} = 0.1$ and vary $v_{\mathrm{s}}$. In both cases, increasing absolute value of either parameter enhances the late-time Integrated Sachs–Wolfe (ISW) effect, leading to excess power at low multipoles. Simultaneously, the early ISW effect and the shift in the angular diameter distance to last scattering modify the acoustic peak structure.

Figure~\ref{fig:Post_P2} presents the one-dimensional posterior distributions and two-dimensional marginalized contours for Model~2, using various combinations of cosmological datasets. The inclusion of low-redshift data such as SDSS, PantheonPlus, and DESI clearly reduces the allowed parameter volume in the $(\Omega_{\mathrm{s}}, v_{\mathrm{s}})$ plane. In particular, the addition of DESI breaks the degeneracies involving $\Omega_{\mathrm{s}}$ and significantly tightens the constraints on $\Omega_c h^2$ and $n_{\mathrm{s}}$. Notably, $v_{\mathrm{s}}$ does not exhibit correlations with the standard cosmological parameters, but only with $\Omega_{\mathrm{s}}$.

%%%%%%%%%%%%%%%%%%%%%%%%%%%%%%%%%%%%%%%%%%%%%%%%%%%%%%%
\begin{table*}[htbp]
\scriptsize
\centering
\resizebox{\textwidth}{!}{%
\begin{tabular}{l|ccccccc}
\toprule
\textbf{Parameters} 
& \textbf{CMB} 
& \textbf{CMB} 
& \textbf{CMB} 
& \textbf{CMB+SDSS}
& \textbf{CMB} 
& \textbf{CMB} 
& \textbf{CMB+DESI} \\
&
& \textbf{+SDSS} 
& \textbf{+PantheonPlus} 
& \textbf{+PantheonPlus}
& \textbf{+DESI} 
& \textbf{+DESI+DESY5} 
& \textbf{+PantheonPlus} \\
\midrule
$\Omega_b\,h^2$ 
& $0.02249 \pm 0.00016$ 
& $0.02239 \pm 0.00014$ 
& $0.02235 \pm 0.00014$ 
& $0.02238 \pm 0.00014$
& $0.02242 \pm 0.00013$ 
& $0.02246 \pm 0.00014$ 
& $0.02245 \pm 0.00014$ \\[6pt]

$\Omega_c\,h^2$ 
& $0.1186 \pm 0.0015$ 
& $0.11993 \pm 0.00098$ 
& $0.1204 \pm 0.0011$ 
& $0.11999 \pm 0.00099$
& $0.11943 \pm 0.00094$ 
& $0.11887 \pm 0.00094$ 
& $0.11910 \pm 0.00094$ \\[6pt]

$100\,\theta_s$ 
& $1.04108 \pm 0.00032$ 
& $1.04093 \pm 0.00030$ 
& $1.04088 \pm 0.00030$ 
& $1.04094 \pm 0.00030$
& $1.04100 \pm 0.00029$ 
& $1.04110 \pm 0.00029$ 
& $1.04106 \pm 0.00030$ \\[6pt]

$\tau_{\mathrm{reio}}$ 
& $0.0504^{+0.0083}_{-0.0072}$ 
& $0.0528 \pm 0.0077$ 
& $0.0540 \pm 0.0076$ 
& $0.0542 \pm 0.0075$
& $0.0537 \pm 0.0076$ 
& $0.0573 \pm 0.0078$ 
& $0.0558 \pm 0.0078$ \\[6pt]

$n_{\mathrm{s}}$ 
& $0.9690 \pm 0.0047$ 
& $0.9656 \pm 0.0039$ 
& $0.9645 \pm 0.0039$ 
& $0.9655 \pm 0.0038$
& $0.9669 \pm 0.0037$ 
& $0.9684 \pm 0.0038$ 
& $0.9678 \pm 0.0037$ \\[6pt]

$\ln\bigl(10^{10}\,A_{\mathrm{s}}\bigr)$ 
& $3.030^{+0.017}_{-0.015}$ 
& $3.040 \pm 0.014$ 
& $3.045 \pm 0.014$ 
& $3.044 \pm 0.014$
& $3.041 \pm 0.014$ 
& $3.049 \pm 0.014$ 
& $3.046 \pm 0.014$ \\[6pt]

$\Omega_{\mathrm{s}}$ 
& $-0.050^{+0.020}_{-0.030}$ 
& $-0.0193 \pm 0.0090$ 
& $0.002 \pm 0.012$ 
& $-0.0106 \pm 0.0083$
& $-0.0207 \pm 0.0081$ 
& $-0.0082 \pm 0.0080$ 
& $-0.0133 \pm 0.0079$ \\[6pt]
\midrule

$H_0$ [km/s/Mpc] 
& $73.4 \pm 3.6$ 
& $69.12 \pm 0.77$ 
& $67.0 \pm 1.1$ 
& $68.29 \pm 0.64$
& $69.48 \pm 0.53$ 
& $68.59 \pm 0.48$ 
& $68.95 \pm 0.49$ \\[6pt]

$\sigma_{8}$ 
& $0.865 \pm 0.031$ 
& $0.8302 \pm 0.0097$ 
& $0.811 \pm 0.012$ 
& $0.8224 \pm 0.0092$
& $0.8309 \pm 0.0098$ 
& $0.8188 \pm 0.0094$ 
& $0.8237 \pm 0.0096$ \\[6pt]

$S_{8}$ 
& $0.811 \pm 0.018$ 
& $0.8292 \pm 0.0096$ 
& $0.837 \pm 0.011$ 
& $0.8314 \pm 0.0097$
& $0.8242 \pm 0.0091$ 
& $0.8212 \pm 0.0092$ 
& $0.8224 \pm 0.0094$ \\[6pt]

$\Omega_{\mathrm{m}}$ 
& $0.265^{+0.024}_{-0.032}$ 
& $0.2994 \pm 0.0072$ 
& $0.320 \pm 0.011$ 
& $0.3067 \pm 0.0062$
& $0.2952 \pm 0.0044$ 
& $0.3018 \pm 0.0041$ 
& $0.2991 \pm 0.0042$ \\[6pt]
\midrule

$\Delta \chi^2_{\mathrm{min},\,\Lambda \mathrm{CDM}}$ 
& $-6.77$ 
& $-2.46$ 
& $0.53$ 
& $-1.56$
& $-4.06$ 
& $-0.61$ 
& $-0.92$ \\[6pt]

$\Delta \ln \mathcal{Z}_{\,\Lambda{\rm CDM}}$ 
& $-0.93$ 
& $-2.31$ 
& $-4.07$ 
& $-3.81$
& $-1.95$ 
& $-4.3$ 
& $-3.59$ \\[6pt]

\bottomrule
\end{tabular}
}
\caption{Observational constraints at 68\% CL and upper limits at 95\% CL obtained for Model~3 using various datasets. Negative values of $\Delta \chi^2_{\mathrm{min},\,\Lambda \mathrm{CDM}}$ favor Model~3 over the standard $\Lambda$CDM scenario, while positive values of $\Delta \ln \mathcal{Z}_{\,\Lambda{\rm CDM}}$ indicate a preference for Model~3.}
\label{tab:model3}
\end{table*}

\begin{figure*}[htbp]
    \centering
    \includegraphics[width=0.9\linewidth]{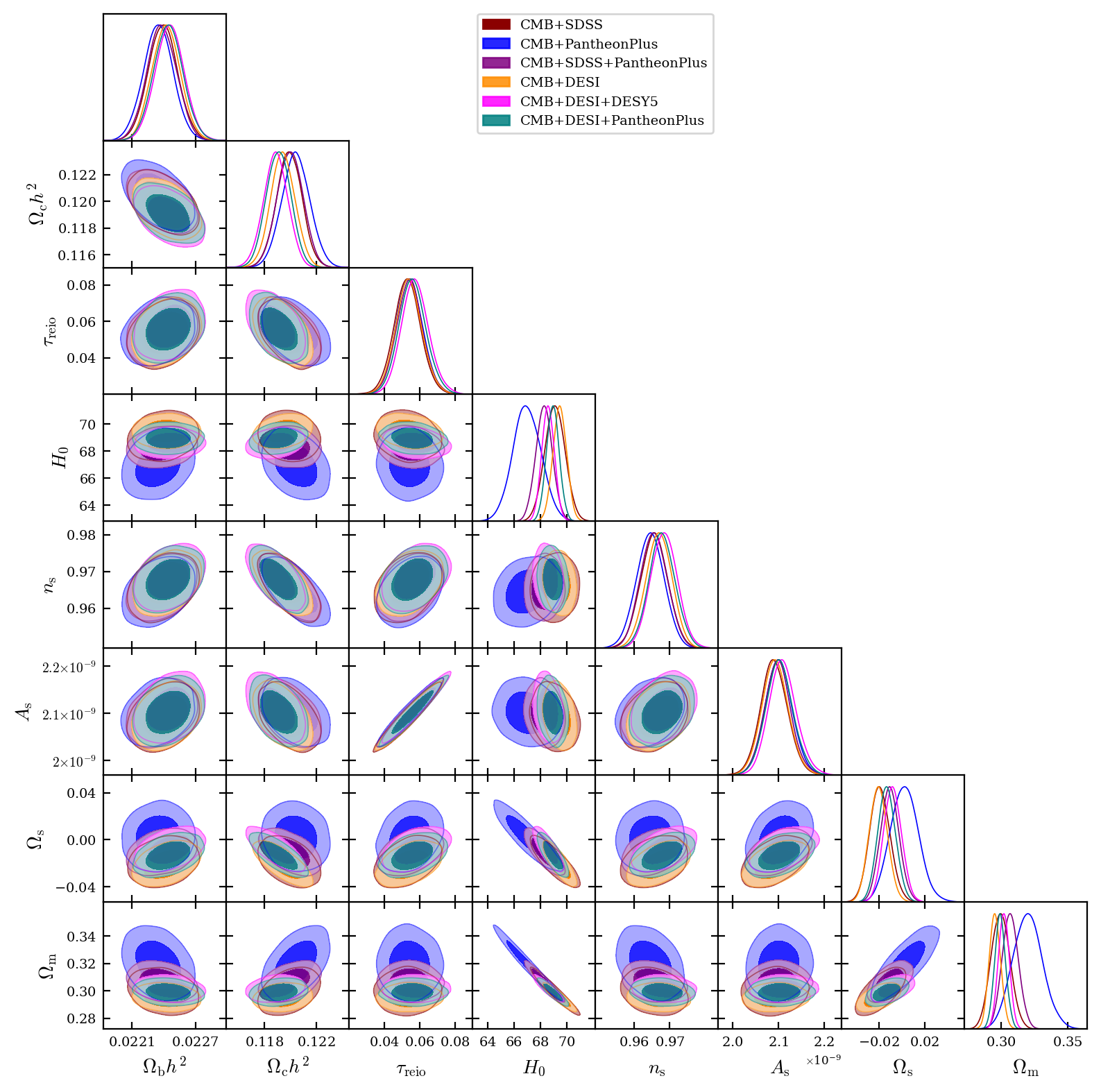}
\caption{One-dimensional posterior distributions and two-dimensional marginalized contours for Model~3, based on various combinations of cosmological datasets. The parameter $\Omega_{\mathrm{s}}$ is allowed to take both positive and negative values, corresponding to a non-relativistic network of cosmic strings. }
    \label{fig:Post_P3}
\end{figure*}

\section{Discussion}\label{discussion}

\subsection{Negative energy density?}

We consider a generalization of the previous scenarios by allowing the effective energy density of the string network, $\Omega_{\mathrm{s}}$, to take both positive and negative values. This setup accommodates a broader class of dynamical dark energy behaviors, including the possibility that the string-like component acts as a negative energy fluid at late times, thereby contributing to an enhanced expansion rate. In this context, the study of negative energy density states may be complemented by cosmological tools, which could indirectly probe violations of standard energy conditions.
To this end, we allow the prior on the cosmic string energy density parameter to span both positive and negative initial values, as reported in the last two columns of table~\ref{tab:priors}. We begin by discussing the results from the run with Model~3, where the string network is assumed to be non-relativistic.

As shown in table~\ref{tab:model3}, the posterior distributions for $\Omega_{\mathrm{s}}$ are significantly shifted toward negative values across all dataset combinations, particularly when CMB data are used alone or in combination with low-redshift measurements, except in the case of CMB+PantheonPlus, where $\Omega_{\mathrm{s}}$ remains fully consistent with zero. For example, \textit{Planck}-only data yield $\Omega_{\mathrm{s}} = -0.050^{+0.020}_{-0.030}$ at 68\% CL, with the negative mean value persisting even after the inclusion of SDSS or DESI data. This trend suggests a mild statistical preference for models in which the effective dark energy component includes a negative-energy contribution beyond that of a cosmological constant.
The inclusion of this negative-energy component impacts several derived cosmological parameters. Most notably, the Hubble constant is significantly increased, thereby reducing the Hubble tension relative to $\Lambda$CDM. With CMB-only data, we find $H_0 = 73.4 \pm 3.6\,\mathrm{km\,s^{-1}\,Mpc^{-1}}$, while the addition of other datasets yields values in the range $H_0 = 68$–$70\,\mathrm{km\,s^{-1}\,Mpc^{-1}}$, with the lowest value observed for CMB+PantheonPlus, $H_0 \sim 67\,\mathrm{km\,s^{-1}\,Mpc^{-1}}$. Similarly, the $S_8$ parameter is slightly lower in the CMB-only analysis, but becomes consistent with low-redshift observations once SDSS, DESI, and supernova datasets are included.

However, although Model~3 leads to improved best-fit likelihoods and a partial alleviation of the Hubble tension, it remains disfavored from a Bayesian perspective due to its additional parameter freedom. The difference in minimum chi-square values relative to $\Lambda$CDM is consistently negative across nearly all dataset combinations, with the largest improvement, $\Delta \chi^2 = -6.77$, achieved for CMB-only data. This indicates that the extended model provides a statistically better fit to the data in those cases.
Nevertheless, the difference in Bayesian evidence is negative for all dataset combinations, suggesting that the improvement in fit is insufficient to overcome the Occam’s razor penalty associated with the extra freedom in $\Omega_{\mathrm{s}}$, which is allowed to vary over both positive and negative values. 
To sum up, while Model~3 is compatible with current data and offers a marginal improvement in the likelihood, it is not statistically preferred over the standard $\Lambda$CDM model.

The parameter constraints for Model~3 are shown in Fig.~\ref{fig:Post_P3}, including the one-dimensional posterior distributions and two-dimensional marginalized contours for $\Omega_{\mathrm{s}}$ and the relevant cosmological parameters. Different dataset combinations highlight their respective roles in constraining the extended parameter space. In particular, the inclusion of large-scale structure data and SNe~Ia significantly narrows the posterior distributions, breaking degeneracies present in the CMB-only case.
The results indicate a preference for slightly negative values of $\Omega_{\mathrm{s}}$ across multiple datasets, consistent with the values reported in table~\ref{tab:model3}. This behavior is also linked to the geometrical degeneracy between $\Omega_{\mathrm{s}}$, $H_0$, and $\Omega_{\mathrm{m}}$, leading to slightly higher Hubble constant values and lower matter density values. However, the overlap with $\Omega_{\mathrm{s}} = 0$ within the $2\sigma$ level confirms the statistical compatibility of the model with $\Lambda$CDM.

\subsection{Results for unconstrained energy and velocity of the string network}

\begin{figure*}[htbp]
    \centering
    \includegraphics[width=0.9\linewidth]{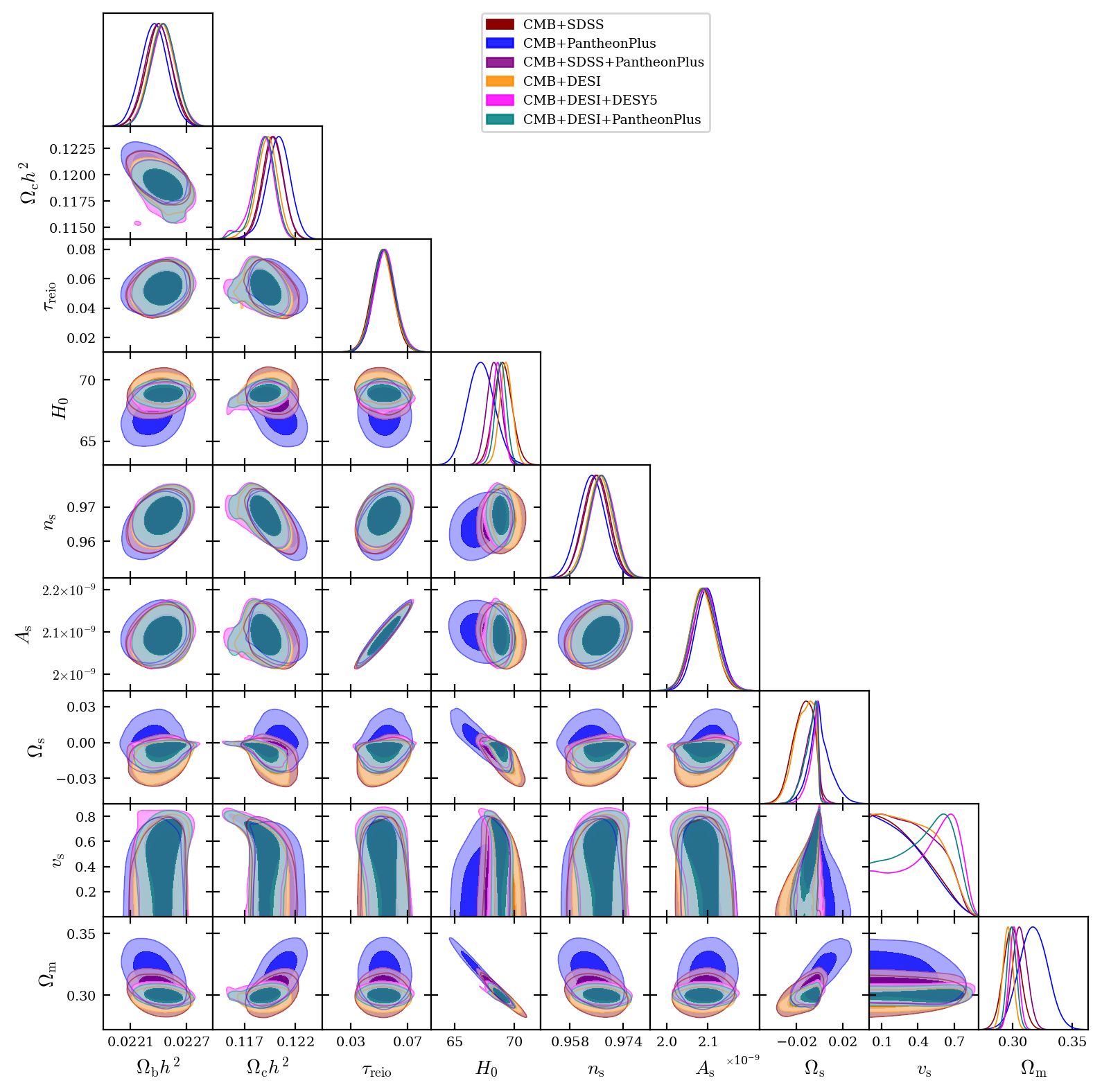}
\caption{One-dimensional posterior distributions and two-dimensional marginalized contours for Model~4, based on various combinations of cosmological datasets. Both the energy density and the bulk velocity of the strings are allowed to vary freely within the ranges defined in table~\ref{tab:priors}. The legend indicates the colors corresponding to different dataset combinations.}
    \label{fig:Post_P4}
\end{figure*}

Model~4 represents the most general scenario considered in this work, allowing both the string energy density parameter and the network velocity to vary freely, as specified in table~\ref{tab:priors}. This extended framework accommodates a broad class of string-inspired dark energy models, including those with negative energy contributions and relativistic dynamics. As shown in table~\ref{tab:model4}, the constraints on $\Omega_{\mathrm{s}}$ continue to favor slightly negative values across all dataset combinations—except for CMB+PantheonPlus, consistent with the findings for Model~3. For example, CMB-only data yield $\Omega_{\mathrm{s}} = -0.038^{+0.029}_{-0.022}$, while the inclusion of DESI and PantheonPlus tightens the bounds to $\Omega_{\mathrm{s}} = -0.0080^{+0.0075}_{-0.0029}$.
The bulk velocity of the string network remains largely unconstrained at the 95\% CL for most dataset combinations, with upper bounds ranging from $v_{\mathrm{s}} < 0.574$ (CMB-only) to $v_{\mathrm{s}} < 0.706$ (CMB+DESI). However, when SN data are added to CMB+DESI, the velocity becomes constrained at 68\% CL, yielding $v_{\mathrm{s}} = 0.48^{+0.32}_{-0.15}$ for CMB+DESI+DESY5 and $v_{\mathrm{s}} = 0.44^{+0.30}_{-0.18}$ for CMB+DESI+PantheonPlus.
These results are consistent with theoretical expectations from string simulations, which predict velocities in the range $v_{\mathrm{s}} \sim 0.3$–$0.6$. However, the data do not show a statistically significant preference for non-zero values, as even in the cases where $v_{\mathrm{s}}$ is bounded at 68\% CL, $v_{\mathrm{s}}$ remains consistent with zero at 95\% CL, as shown in figure~\ref{fig:Post_P4}.
Model comparison indicates a consistent improvement in fit relative to $\Lambda$CDM, with $\Delta \chi^2$ reaching values as low as $-6.58$ for the CMB+DESI combination. Nonetheless, the Bayesian evidence remains negative across all dataset combinations, with values ranging from $-1.74$ to $-4.75$, implying that the data do not justify the additional parameter complexity of this extended model.

The inclusion of the string network velocity does not significantly affect the other parameters of the model. For example, in the CMB-only case, the Hubble constant remains consistent with the local value, with $H_0 = 72.7^{+3.0}_{-3.7}\,\mathrm{km\,s^{-1}\,Mpc^{-1}}$ at 68\% CL. The addition of low-redshift data shifts this estimate toward lower values, around $68$–$69\,\mathrm{km\,s^{-1}\,Mpc^{-1}}$. As a consequence, the matter density is relatively low in the CMB-only case, with $\Omega_{\mathrm{m}} = 0.270 \pm 0.027$ at 68\% CL, but converges toward the $\Lambda$CDM value when low-redshift data are included.

Figure~\ref{fig:Post_P4} presents the one-dimensional posterior distributions and two-dimensional marginalized contours for Model~4, obtained using various combinations of cosmological datasets. The figure illustrates the effects of allowing both the string energy density and the bulk velocity to vary freely. As in Model~3, the posteriors for $\Omega_{\mathrm{s}}$ show a mild preference for negative values across all dataset combinations, with tighter constraints obtained when low-redshift data are included. The contours reveal parameter degeneracies, particularly between $\Omega_{\mathrm{s}}$ and $H_0$, as well as between $v_{\mathrm{s}}$ and $\Omega_c h^2$, which become more pronounced with the inclusion of additional datasets. Despite these correlations, the string network dynamics remain weakly constrained, as the posterior distribution for $v_{\mathrm{s}}$ remains broad, limiting the statistical significance of any detection.

%%%%%%%%%%%%%%%%%%%%%%%%%%%%%%%%%%%%%%%%%%%%%%%%%%%%%%%
\begin{table*}[htbp]
\scriptsize
\centering
\resizebox{\textwidth}{!}{%
\begin{tabular}{l|ccccccc}
\toprule
\textbf{Parameters} 
& \textbf{CMB} 
& \textbf{CMB} 
& \textbf{CMB} 
& \textbf{CMB+SDSS}
& \textbf{CMB} 
& \textbf{CMB}
& \textbf{CMB+DESI} \\
&
& \textbf{+SDSS} 
& \textbf{+PantheonPlus} 
& \textbf{+PantheonPlus}
& \textbf{+DESI} 
& \textbf{+DESI+DESY5}
& \textbf{+PantheonPlus} \\
\midrule
$\Omega_b\,h^2$ 
& $0.02250 \pm 0.00017$ 
& $0.02241 \pm 0.00014$ 
& $0.02236 \pm 0.00014$ 
& $0.02240 \pm 0.00014$
& $0.02244 \pm 0.00014$ 
& $0.02245 \pm 0.00014$
& $0.02245 \pm 0.00013$ \\[6pt]

$\Omega_c\,h^2$ 
& $0.1185 \pm 0.0015$ 
& $0.1197 \pm 0.0011$ 
& $0.1203 \pm 0.0012$ 
& $0.1198 \pm 0.0010$
& $0.1192^{+0.0011}_{-0.00092}$ 
& $0.1187^{+0.0012}_{-0.00086}$
& $0.1189^{+0.0011}_{-0.00087}$ \\[6pt]

$100\,\theta_s$ 
& $1.04109 \pm 0.00033$ 
& $1.04096 \pm 0.00032$ 
& $1.04089 \pm 0.00031$ 
& $1.04096 \pm 0.00030$
& $1.04105^{+0.00029}_{-0.00033}$ 
& $1.04115^{+0.00028}_{-0.00038}$
& $1.04111^{+0.00028}_{-0.00034}$ \\[6pt]

$\tau_{\mathrm{reio}}$ 
& $0.0503 \pm 0.0083$ 
& $0.0524 \pm 0.0078$ 
& $0.0537 \pm 0.0075$ 
& $0.0534 \pm 0.0076$
& $0.0527 \pm 0.0078$ 
& $0.0541 \pm 0.0080$
& $0.0534 \pm 0.0079$ \\[6pt]

$n_{\mathrm{s}}$ 
& $0.9691 \pm 0.0047$ 
& $0.9660 \pm 0.0038$ 
& $0.9646 \pm 0.0040$ 
& $0.9657 \pm 0.0038$
& $0.9671 \pm 0.0037$ 
& $0.9677 \pm 0.0037$
& $0.9675 \pm 0.0037$ \\[6pt]

$\ln\bigl(10^{10}\,A_{\mathrm{s}}\bigr)$ 
& $3.030 \pm 0.017$ 
& $3.039 \pm 0.015$ 
& $3.044 \pm 0.014$ 
& $3.042 \pm 0.014$
& $3.039 \pm 0.015$ 
& $3.042 \pm 0.015$
& $3.040 \pm 0.015$ \\[6pt]

$\Omega_{\mathrm{s}}$ 
& $-0.038^{+0.029}_{-0.022}$ 
& $-0.0141^{+0.011}_{-0.0068}$ 
& $0.0005^{+0.0081}_{-0.0098}$ 
& $-0.0074^{+0.0072}_{-0.0045}$
& $-0.0135^{+0.011}_{-0.0054}$ 
& $-0.0054^{+0.0053}_{-0.0025}$
& $-0.0080^{+0.0075}_{-0.0029}$ \\[6pt]

$v_{\mathrm{s}}$ 
& $< 0.574$ 
& $< 0.657$ 
& $< 0.656$ 
& $< 0.705$
& $< 0.706$ 
& $0.48^{+0.32}_{-0.15}$
& $0.44^{+0.30}_{-0.18}$ \\[6pt]
\midrule

$H_0$ [km/s/Mpc] 
& $72.7^{+3.0}_{-3.7}$ 
& $69.05 \pm 0.77$ 
& $67.2 \pm 1.1$ 
& $68.31 \pm 0.62$
& $69.31 \pm 0.53$ 
& $68.60^{+0.49}_{-0.42}$
& $68.87 \pm 0.48$ \\[6pt]

$\sigma_{8}$ 
& $0.860^{+0.027}_{-0.031}$ 
& $0.8302 \pm 0.0098$ 
& $0.813 \pm 0.013$ 
& $0.8236 \pm 0.0092$
& $0.8311 \pm 0.0095$ 
& $0.8246 \pm 0.0094$
& $0.8270 \pm 0.0092$ \\[6pt]

$S_{8}$ 
& $0.813 \pm 0.017$ 
& $0.8295 \pm 0.0099$ 
& $0.837 \pm 0.011$ 
& $0.8321 \pm 0.0098$
& $0.8258 \pm 0.0093$ 
& $0.8266 \pm 0.0097$
& $0.8262 \pm 0.0094$ \\[6pt]

$\Omega_{\mathrm{m}}$ 
& $0.270 \pm 0.027$ 
& $0.2996 \pm 0.0072$ 
& $0.318 \pm 0.012$ 
& $0.3062 \pm 0.0062$
& $0.2962 \pm 0.0042$ 
& $0.3015 \pm 0.0039$
& $0.2995 \pm 0.0039$ \\[6pt]
\midrule

$\Delta \chi^2_{\mathrm{min},\,\Lambda \mathrm{CDM}}$ 
& $-6.07$ 
& $-2.95$ 
& $0.36$ 
& $-2.44$
& $-6.58$ 
& $-5.47$
& $-5.99$ \\[6pt]

$\Delta \ln \mathcal{Z}_{\,\Lambda{\rm CDM}}$ 
& $-1.74$ 
& $-2.94$ 
& $-4.75$ 
& $-4.12$
& $-2.18$ 
& $-3.88$
& $-3.06$ \\[6pt]

\bottomrule
\end{tabular}
}
\caption{Observational constraints at 68\% CL and upper limits at 95\% CL obtained for Model~4 using various datasets. Negative values of $\Delta \chi^2_{\mathrm{min},\,\Lambda \mathrm{CDM}}$ favor Model~4 over the standard $\Lambda$CDM scenario, while positive values of $\Delta \ln \mathcal{Z}_{\,\Lambda \mathrm{CDM}}$ indicate a preference for Model~4.}
\label{tab:model4}
\end{table*}

\section{Conclusions}\label{conclusions}

The four phenomenological models explored in this work represent progressively more general extensions of the standard $\Lambda$CDM framework, incorporating a cosmic string network with increasing degrees of freedom. Model~1, featuring a positive energy density and non-relativistic scaling, constitutes a minimal extension and yields tight upper bounds on the string contribution, with $\Omega_{\mathrm{s}} < 0.00824$ at 95\% CL from the CMB+DESI dataset. Model~2 introduces velocity dependence, resulting in $\Omega_{\mathrm{s}} < 0.00663$ and $v_{\mathrm{s}} < 0.592$ for CMB+DESI at 95\% CL, but shows no preference for relativistic string dynamics. In both cases, the best-fit values of $H_0$ and $S_8$ shift slightly in the direction of alleviating known cosmological tensions, although the changes are insufficient, and neither model is favored from a Bayesian perspective. Model~3 allows the string energy density to take both positive and negative values, enabling a broader class of exotic energy contributions. Across all datasets, we find a consistent preference for slightly negative values of $\Omega_{\mathrm{s}}$, with CMB-only data yielding $\Omega_{\mathrm{s}} = -0.050^{+0.020}_{-0.030}$ and an improvement in the best-fit likelihood of $\Delta \chi^2 = -6.77$. However, this improved fit comes at the cost of increased model complexity, resulting in negative Bayesian evidence.
Model~4 generalizes this further by allowing both $\Omega_{\mathrm{s}}$ and $v_{\mathrm{s}}$ to vary freely. The most stringent constraints are obtained for the combination CMB+DESI+DESY5, with $\Omega_{\mathrm{s}} = -0.0054^{+0.0053}_{-0.0025}$ and $v_{\mathrm{s}} = 0.48^{+0.32}_{-0.15}$ at 68\% CL. While the data continue to favor mildly negative energy densities and a broad range of string velocities, the Bayesian evidence does not support the added complexity of the model.
In summary, although cosmic string-inspired extensions of $\Lambda$CDM can yield improved fits and partially alleviate cosmological tensions, there is currently no strong statistical evidence favoring their inclusion over the standard model. Future CMB and large-scale structure surveys with improved sensitivity to late-time dynamics may offer sharper insights into the viability of such exotic energy components.

\acknowledgments
HC acknowledges the support of the China Scholarship Council (CSC) program (Project ID No.\ 202406230341). HC and LV acknowledge support by the National Natural Science Foundation of China (NSFC) through the grant No.\ 12350610240 ``Astrophysical Axion Laboratories''. EDV is supported by a Royal Society Dorothy Hodgkin Research Fellowship. LV also thanks Istituto Nazionale di Fisica Nucleare (INFN) through the ``QGSKY'' Iniziativa Specifica project. This article is based upon work from COST Action CA21136 ``Addressing observational tensions in cosmology with systematics and fundamental physics'' (CosmoVerse), supported by COST (European Cooperation in Science and Technology). The authors acknowledge the use of High-Performance Computing resources from the IT Services at the University of Sheffield.

%++++++++++++++++++++++++++++++++
\bibliographystyle{JHEP}
\bibliography{biblio}
%++++++++++++++++++++++++++++++++
\end{document}